\documentclass[reprint,aps,pra,showpacs,amsmath,amssymb,superscriptaddress]{revtex4-1}

\usepackage{xcolor}
\usepackage{graphicx}
\usepackage{placeins}
\usepackage[caption=false]{subfig} 

\usepackage{braket}
\usepackage{comment}

\newcommand{\br}{{\bf r}}

\begin{document}

\title{Quantum phases of disordered three-dimensional Majorana-Weyl fermions}

\author{Justin H. Wilson}
\affiliation{Institute of Quantum Information and Matter and Department of Physics, California Institute of Technology, Pasadena, CA 91125 USA}
  \affiliation{Condensed Matter Theory Center and Joint Quantum Institute, Department of Physics, University of Maryland, College Park, Maryland 20742-4111, USA}

 \author{J. H. Pixley}
  \affiliation{Condensed Matter Theory Center and Joint Quantum Institute, Department of Physics, University of Maryland, College Park, Maryland 20742-4111, USA} 

 \author{Pallab Goswami}
  \affiliation{Condensed Matter Theory Center and Joint Quantum Institute, Department of Physics, University of Maryland, College Park, Maryland 20742-4111, USA} 

 \author{S. Das Sarma}
\affiliation{Condensed Matter Theory Center and Joint Quantum Institute, Department of Physics, University of Maryland, College Park, Maryland 20742-4111, USA} 

\begin{abstract}
The gapless Bogoliubov-de Gennes (BdG) quasiparticles of a clean three dimensional spinless $p_x+ip_y$ superconductor provide an intriguing example of a thermal Hall semimetal (ThSM) phase of Majorana-Weyl fermions; such a phase can support a large anomalous thermal Hall conductivity and protected surface Majorana-Fermi arcs at zero energy. 
We study the effects of quenched disorder on such a gapless topological phase by carrying out extensive numerical and analytical calculations on a lattice model for a disordered, spinless $p_x+ip_y$ superconductor. Using the kernel polynomial method, we compute both average and typical density of states for the BdG  quasiparticles, from which we construct the phase diagram of three dimensional dirty $p_x+ip_y$ superconductors as a function of disorder strength and chemical potential of the underlying normal state. We establish that the power law quasi-localized states induced by rare statistical fluctuations of the disorder potential give rise to an exponentially small density of states at zero energy, and even infinitesimally weak disorder converts the ThSM into a thermal diffusive Hall metal (ThDM). Consequently, the phase diagram of the disordered model only consists of ThDM and thermal insulating phases. We show the existence of two types of thermal insulators: (i) a trivial thermal band insulator (ThBI) [or BEC phase] with a smeared gap that can occur for suitable band parameters and all strengths of disorder, supporting only exponentially localized Lifshitz states (at low energy), and (ii) a thermal Anderson insulator that only exists for large disorder strengths compared to all band parameters. We determine the nature of the two distinct localization-delocalization transitions between these two types of insulators and ThDM. Additionally, we establish the scaling properties of an avoided (or hidden) quantum critical point for moderate disorder strengths, which govern the crossover between ThSM and ThDM phases over a wide range of energy scales. We also discuss the experimental relevance of our findings for three dimensional, time reversal symmetry breaking, triplet superconducting states.  
\end{abstract}

\pacs{%
} %Insert PACS

\maketitle

\section{Introduction}
\label{sec:intro}
Three dimensional Weyl semimetals (WSM) are one the most prominent examples of a gapless topological phase; two non-degenerate bands touch at isolated points in the Brillouin zone causing the low energy quasiparticles to exhibit a linear dispersion in the vicinity of these diabolic points~\cite{Teo2016}. Such band touching points act as the (anti)monopoles of Abelian Berry curvature, giving rise to protected surface Fermi arcs, and exotic transport and electrodynamic properties~\cite{Zyuzin3,Grushin,GoswamiTewari,Vazifeh,Aji,Son,Vishwanathsid,Goswamianomaly,Gyrotropy1}. Since the momentum separation vector $\delta \mathbf{K}$ between the right and left handed Weyl points selects out a preferred inertial frame, a WSM phase breaks Lorentz invariance despite possessing a linear dispersion~\cite{Zyuzin3,Grushin,GoswamiTewari}.  

Interestingly, the low energy quasiparticles of several time-reversal ($\mathcal{T}$) and inversion ($\mathcal{I}$) symmetry breaking ground states of strongly correlated materials can be described by Weyl fermions, which may or may not possess a conserved electric charge. The gapless quasiparticles of various magnetically ordered states of 227 pyrochlore iridates~\cite{Vishwanath2011,Balents2014,Goswami-iridates}, and time reversal symmetry breaking $p_x+ip_y$ ~\cite{Volovik,Meng,Gong,SauTewari} and $d_{xz}+id_{yz}$ ~\cite{GoswamiBalicas,GoswamiNevidomskyy} superconductors are some examples of proposed Weyl excitations in correlated materials. According to the Altland-Zirnbauer classification scheme~\cite{AltlandZirnbauer} of noninteracting fermions, a $\mathcal{T}$ breaking WSM can be a member of the following three symmetry classes: unitary/A (a magnetic phase with conserved electric charge), C (chiral, spin singlet pairing like $d_{xz}+id_{yz}$), and D (spin rotational symmetry breaking chiral paired states like $p_x+ip_y$). While class A WSMs can exhibit both anomalous charge and thermal Hall effects, a class D WSM only exhibits an anomalous thermal Hall effect. By contrast, a class C WSM can support both anomalous spin (in response to a spin electric field caused by a spatially varying Zeeman coupling) and thermal Hall effects~\cite{GoswamiBalicas}. We also note that $\mathcal{T}$ preserving but $\mathcal{I}$ breaking WSMs  , which possess an even number of left and right handed nodal points can be realized for symmetry classes AII (spin rotation breaking with conserved electric charge) and DIII (spin rotation breaking superconductors). The experimentally observed WSMs in spin orbit coupled, noncentrosymmetric materials~\cite{Weng,HuangHasan,XuHasan,Lv,XuHasan1} are members of the symmetry class AII. Such WSMs do not support any anomalous Hall effect, but they can exhibit natural optical activity~\cite{Gyrotropy1}.

Since impurities are ubiquitous in solid state materials, it is extremely important to study disorder effects on the stability of gapless topological states. 
In all realistic situations,  the fate of a disordered WSM is crucial for understanding experimental behavior since disorder is invariably present in all material systems.
The unconventional behavior of disordered Weyl fermions was first discussed by Fradkin~\cite{Fradkin1986,Fradkin1986a}, long before any topological properties of WSM were understood. 
Fradkin's conclusion was that the WSM is stable to weak finite disorder.
Over the past few years, disorder effects on WSMs belonging to class A and class AII have been extensively studied using both analytical~\cite{Shindou2009,Goswami2011,Ominato2014,Bitan-2014,Nandkishore2014,Leo-2015,Syzranov2015,Altland-2015,Syzranov2016,Louvet-2016} and numerical methods~\cite{Kobayashi2014,Sbierski2014,Pixley2015-prl,Sbierski2015,Pixley2015,LiuShindou,Bera2016,Shapourian2015,Pixley2016,Pixley2016a,Sbierski2016a}. Even though some theoretical works have discussed the effects of quenched disorder on three dimensional, gapped topological superconductors~\cite{RyuNomura,GoswamiChakravarty2016}, the problem of dirty Weyl superconductors~\cite{GoswamiNevidomskyy} has remained relatively unexplored. 
In addition, class D superconductors have been studied in detail in both one and two dimensions~\cite{Cho1997,Senthil1998,*Senthil1999,*Senthil1999a,Bocquet2000,Mildenberger2006,Mildenberger2007,Kagalovsky2008,Beenakker2015}, whereas a systematic study in three dimensions remains (to the best of our knowledge) largely unexplored. One aim of the present manuscript is to fill this gap. 
In the present work, we study the effects of quenched disorder on a three dimensional $p_x+ip_y$ superconductor. Interestingly, the Bogoliubov-de Gennes (BdG) Weyl excitations of this phase are real fermions and therefore are in fact Majorana-Weyl (MW) fermions, which can be realized in $^3$He-A~\cite{Volovik} and in time reversal symmetry breaking, triplet paired states of ferromagnetic superconductors~\cite{SauTewari}. Since a $p_x+ip_y$ superconductor can support a large anomalous thermal Hall effect, we will refer to the MW semimetal as a thermal Hall semimetal (ThSM).     

For a comprehensive analysis of dirty superconductors one has to account for the complicated interplay between interaction and disorder. Since unconventional superconductors with finite angular momentum pairing are not protected by Anderson's theorem (which applies only to $s$-wave pairing), non-magnetic disorder can suppress the superconducting transition temperature ($T_c$) of $p$-wave superconductors. This type of disorder induced competition between the normal state (a diffusive Fermi liquid) and an unconventional paired state has been studied for a long time~\cite{Sigrist1991}, and is not the focus of the present work. Rather, we will develop a qualitative understanding of disorder effects on the gapless BdG quasiparticles. Therefore, instead of performing any self-consistent calculation of the pairing gap, we will assume a constant pairing amplitude and study the effects of randomness on a class D quadratic Hamiltonian of a spinless $p_x+ip_y$ three-dimensional superconductor. This should be a reasonable approximation deep inside the superconducting phase, and our interest is understanding the quantum phases of the system rather than the temperature-induced `classical' phase transition.

We begin by describing and summarizing our main findings for the disordered quantum phase diagram of three-dimensional class D MW fermions in Sec.~\ref{sec:results}, which provides a helpful guide for the detailed results derived in the rest of the paper. The remainder of the paper is organized as follows: In Section~\ref{sec:model} we introduce the model of interest including the phases within the clean limit and the numerical method used. 
In Section~\ref{sec:non-pert} we discuss the non-perturbative effects of weak disorder, notably the effect of rare regions.
In Section~\ref{sec:phase} and~\ref{sec:qpt} we discuss, respectively, the phase diagram and the properties of the avoided quantum critical point. 
In Section~\ref{sec:thsm-thbi} we look into the quantum critical line separating the insulating phase from the thermally metallic phase, in Section~\ref{sec:localization} the localization physics is considered up to large disorder, and we conclude in Section~\ref{sec:conclusions}. In
 Appendix~\ref{sec:an}, we present perturbative analytical results for the nature of the quantum phase transitions, in
 Appendix~\ref{sec:ThermalDiffMetal} we study the anti-localization peak in the DOS, and in Appendix~\ref{sec:AppendixA} we discuss the details of determining the phase boundaries.

\section{Main Results}
\label{sec:results}

\begin{figure}[!ht]
\includegraphics[width=0.95\columnwidth]{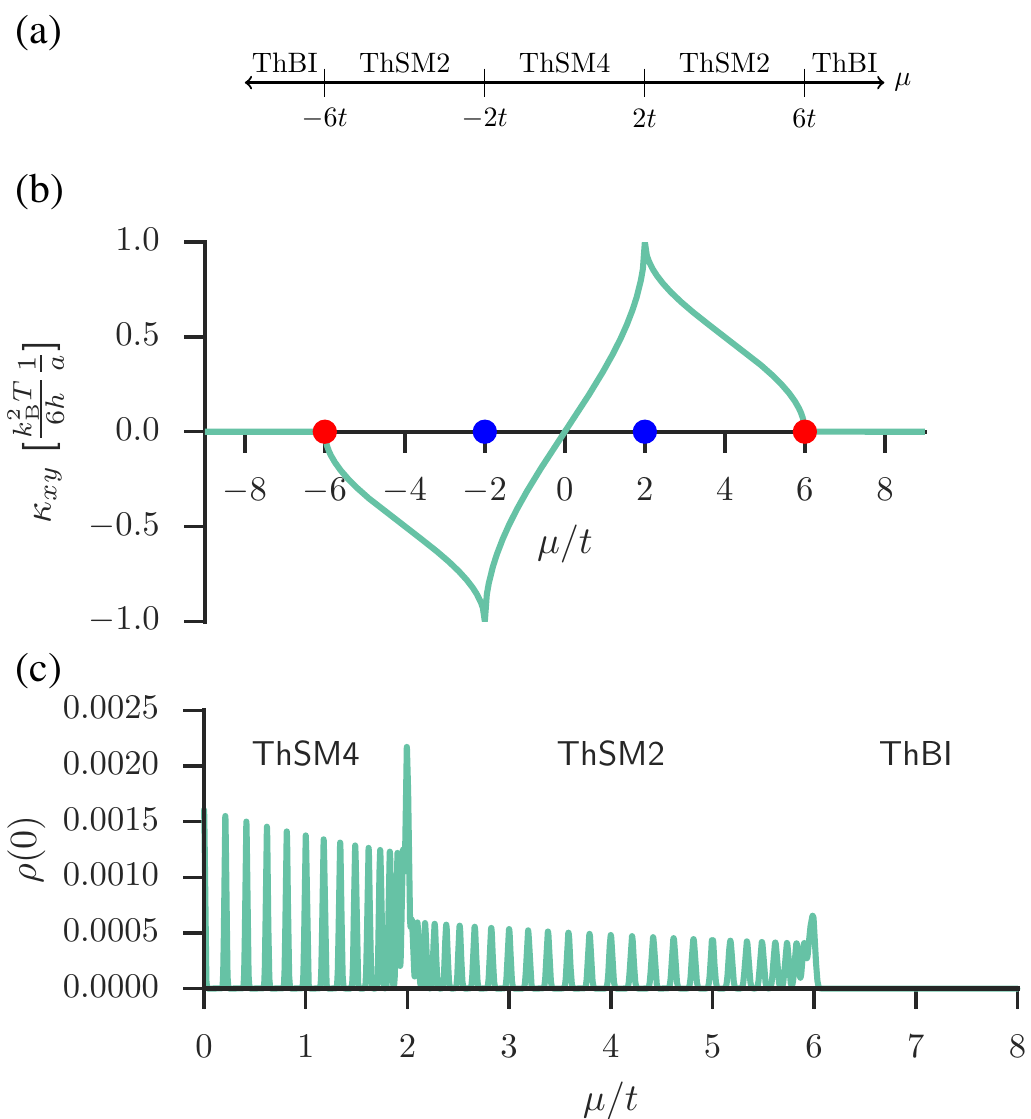}
\caption{(a) Phase diagram of the clean, lattice model for a three dimensional $p_x+ip_y$ superconductor [see Eq.~(\ref{eq:pipHamiltonian})], as a function of the ratio $\mu/t$, where $\mu$ and $t$ are respectively the chemical potential and the nearest neighbor hopping strength for the underlying normal state. Thermal Hall semimetals with two and four Weyl nodes are respectively denoted as ThSM2 and ThSM4, and ThBIs represent fully gapped, thermal insulators arising due to BEC/molecular pairing in the absence of an underlying Fermi surface. 
(b)
Berry curvature induced low temperature thermal Hall conductivity $\kappa_{xy}$ for the clean model. 
The maximum value of $\kappa_{xy}$ is $k^2_B T/(6ha)$, where $a$ is the lattice spacing. This maximum value occurs when two Weyl nodes get shifted to the Brillouin zone boundary. Notice that one chiral, surface Majorana mode contributes $k^2_B T/(6h)$ to $\kappa_{xy}$, and the displayed bulk $\kappa_{xy}$ equals the net contribution from all chiral modes, reflecting the bulk-boundary correspondence. 
(c) Zero energy density of states $\rho(0)$ for a clean ($W=0$), system with linear dimension $L=60$ (in unites of the lattice spacing). We can clearly discern each phase even though in the thermodynamic limit ($L\rightarrow\infty$), $\rho(0) = 0$ in all of these phases.
  The $\delta$-functions in the DOS are broadened by a Gaussian of width $\sigma =  D\pi/2^{11}$ for a bandwidth $D$ (Ref.~\onlinecite{Weisse2006}). 
}
\label{fig:cleanPhaseDiagram}
\end{figure}

Our numerical calculations are performed on a lattice model for a spinless $p_x+ip_y$ three-dimensional superconductor defined on a simple cubic lattice [see Eq.~(\ref{eq:pipHamiltonian})]. In the absence of disorder, the lattice model can support (as a function of chemical potential) three distinct ThSMs (with different numbers of Weyl nodes or diabolic points) and two topologically trivial gapped insulating phases, as shown in Fig.~\ref{fig:cleanPhaseDiagram}. These `quantum phases' arise simply from the `band structure' properties of the BdG quasiparticles as described in depth in Sec.~\ref{sec:model}.
The ThSMs with two and four Weyl points are respectively denoted as ThSM2 and ThSM4, which both carry a non-quantized thermal Hall conductivity. For the clean problem, the density of states (DOS) for MW fermions vanishes as $\rho(E) \sim |E|^2$, while the DOS for a thermal band insulator (ThBI) or BEC phase has a hard spectral gap.
These phases are separated by clean, quantum critical points (QCPs) located at $\mu/t= \pm 2, \; \pm 6$ (for a chemical potential $\mu$ and hopping $t$),
with critical excitations that display anisotropic dispersions $E_{AW}(\mathbf{k})=\pm \sqrt{t^2 k^4_3+ \Delta^2_p k^2_\perp}$ [with $k_\perp=\sqrt{k^2_1+k^2_2}$, and $\Delta_p$ being the pairing amplitude]. Therefore the DOS for such critical excitations vary as $\rho(E) \sim |E|^{3/2}$, and we can associate an effective dynamic scaling exponent from $\rho(E)\sim|E|^{d/z_{aw}-1}$, with $z_{aw}=6/5$. Thus, at the clean anisotropic points the DOS is a non-analytic function of energy (or chemical potential).

The DOS in each region of the phase diagram in Fig.~\ref{fig:cleanPhaseDiagram} (a) vanishes faster than $|E|$ and as a result
disorder acts as an irrelevant perturbation, which suggests that the clean phase diagram is robust to a weak amount of disorder.
(This perturbative argument is similar to what happens in a normal  non-superconducting WSM for weak disorder with the tentative conclusion that the WSM phase is stable to weak disorder.)
One interesting aspect of the present class D model is the possibility for the density of states to become non-analytic (as a predicted consequence of the perturbative RG) since rigorous bounds~\cite{Wegner-1981} do not apply.
The perturbative analysis also predicts that at finite disorder the semimetal and the band insulator have to undergo quantum phase transitions to a diffusive thermal Hall metal (ThDM), possessing a finite density of states at zero energy. 

\begin{figure}[!ht]
  \centering
 \includegraphics[width=\columnwidth]{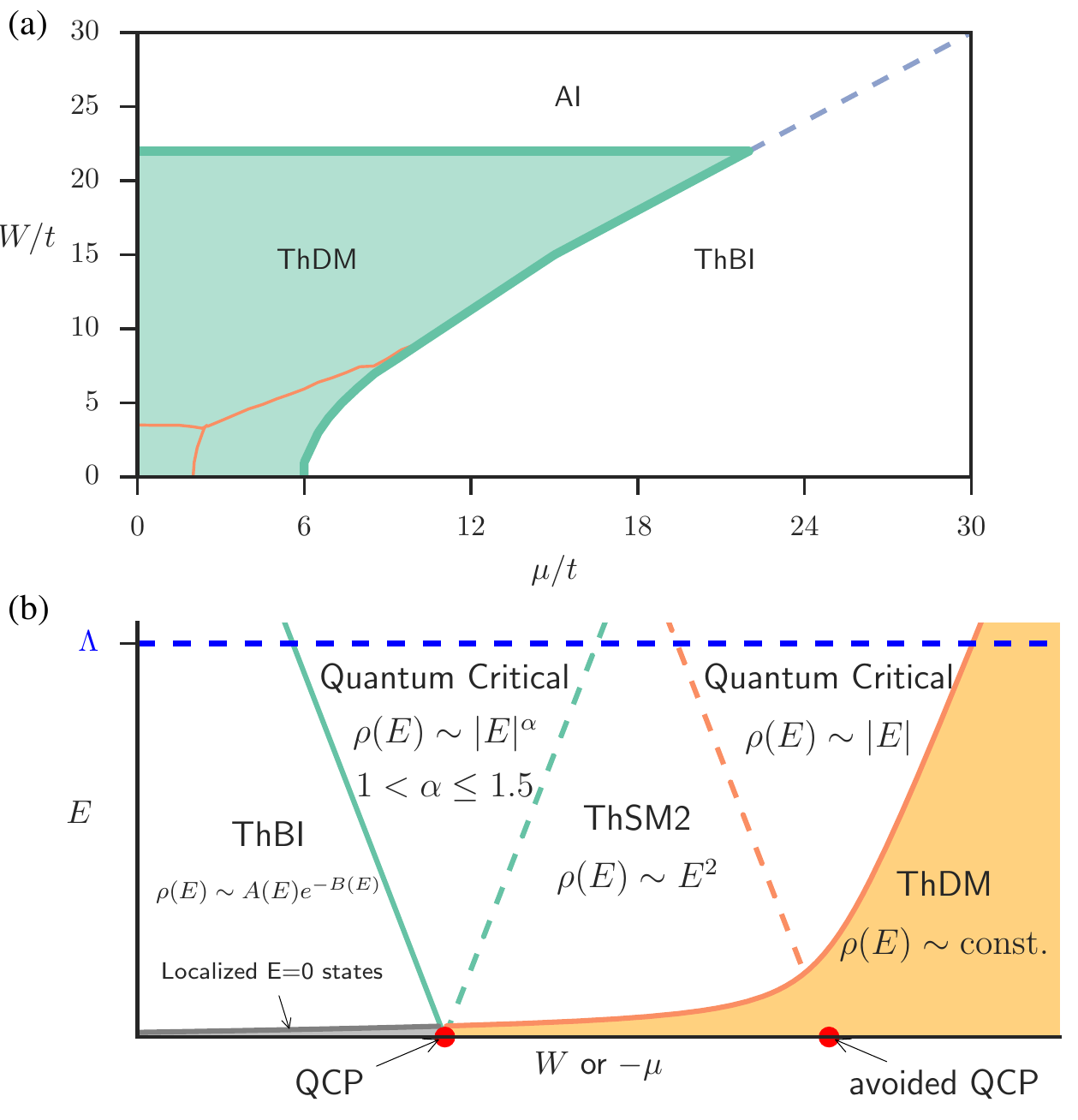} 
  \caption{(a) The zero energy (or temperature) schematic phase diagram of Hamiltonian Eq.~(\ref{eq:pipHamiltonian}) with $\mu$ as the chemical potential for the electrons, $W$ is the strength of disorder, and $t$ is the hopping strength. ThDM is a diffusive thermal Hall metal with finite density of states at zero energy, ThBI is the thermal band insulator supporting exponentially localized Lifshitz states, and AI denotes an Anderson insulator phase of class D, BdG quasiparticles. 
  The thin solid orange line indicates the crossover from thermal Hall semimetals (ThSMs) and ThDM which occur at finite energies, and are governed by an avoided QCP [as shown in (b)]. The dashed line at large $W/t$ and $\mu/t$ represents the possible separation between the two Anderson insulating phases. We expect that a ThDM regime (of a width on the order of the hopping strength) could exist along this line due to the different nature of the two localized phases and their respective transitions.
(b) A sketch of what each regime means [taken as a cut in (a) for either fixed $\mu$ or $W$ such that one passes through \emph{both} the ThBI to ThDM localization transition and the avoided transition], which we verify in each relevant section below.
 Due to the exponentially small DOS at $E=0$ there is only a ThDM at low energies; despite this the DOS still resembles that of a ThSM above a cross over energy scale. The perturbatively predicted  critical point between ThSM and ThDM becomes avoided, but its quantum critical fan can be probed numerically over a wide range of energies.
    The ThBI phase is insulating and an actual critical point separates it from the ThDM phase with its own critical fan. Nonetheless, $\rho(0)>0$ in the ThBI phase purely from localized rare states (represented by the grayed region in the ThBI phase).
    $\Lambda$ is some characteristic energy scale, above which short distance or lattice effects are important.
   This figure does not depict the Anderson localization transition occurring at a very large disorder strength (or energy).
}
  \label{fig:phasediagram}
\end{figure}

However, for very weak disorder we find that non-perturbative effects give rise to quasilocalized rare states that convert the ThSM into a ThDM, with an exponentially small density of states (DOS) at zero energy~\cite{Nandkishore2014,Pixley2016,Pixley2016a} independent of how weak the disorder is. 
By contrast, we find that the gapped ThBI is robust to a weak amount of disorder with a clear average gap in the DOS, and the low energy spectrum is composed of Anderson localized mid-gap states (i.e. Lifshitz states).
Thus, non-perturbative rare region physics destroys the ThSM phase (converting it generically into a ThDM phase even for weak disorder), but does not destroy the ThBI phase (despite Lifshitz states contributing a non-zero DOS inside the average band gap).
Our numerical results are consistent with $\rho(0) \neq 0$ across the entire disordered phase diagram and the DOS is always analytic at low-$|E|$.
Therefore, the numerically determined phase diagram of the disordered model illustrated in Fig.~\ref{fig:phasediagram} (a), only consists of a delocalized ThDM phase, a ThBI (or a trivial BEC) phase supporting exponentially localized Lifshitz states at $E=0$, and a class D Anderson insulator (AI) for very large disorder. The weak disorder controlled transitions and crossovers can be tracked by the average DOS, while an Anderson localization of the ThDM can only be tracked with non-self averaging quantities such as the typical DOS, which we compute with the kernel polynomial method~\cite{Weisse2006} (KPM).
We stress that all of the disorder controlled phases described here are superconductors and therefore exhibit the Meissner effect. However, the ThDM and an ordinary Fermi liquid both show similar linear-in-temperature specific heat
and longitudinal thermal conductivity.

For weak to moderate disorder strengths, the phase diagram exhibits crossover between ThSMs and ThDM over a wide range of energies, governed by an avoided QCP [see Fig.~\ref{fig:phasediagram}(b)]. Our analytical one loop renormalization group (RG) calculations predict that the universality class of the avoided QCP is controlled by the repulsive fixed point of ``axial chemical potential" type disorder for MW fermions, with a dynamic scaling exponent $z=3/2$ and a correlation length exponent $\nu=1$, which agrees well with our numerical calculations. We emphasize that the axial chemical potential disorder is not present in our bare microscopic model, and we are only making a statement about the fixed point Hamiltonian that determines the crossover properties at finite energy between the ThSM and the ThDM. 
The ``axial chemical potential" universality class 
provides a possible explanation for 
why the nonperturbative rare states of the current particle-hole symmetric problem and the rare states of a Dirac (or Weyl) semimetal in the presence of particle-hole asymmetric scalar random potential show similar properties~\cite{Nandkishore2014,Pixley2016,Pixley2016a}. 

We find a genuine localization-delocalization quantum phase transition between the ThBI and ThDM phases. 
Physically, this is due to the average band gap being ``filled in'' by random mid-gap states that have wave functions which overlap significantly with states on neighboring sites resulting in a delocalized phase.
For  infinitesimally weak disorder, perturbing about the anisotropic critical point gives rise to a phase boundary separating the ThDM and ThBI phases. Along this line, perturbatively we find that the DOS still obeys $\rho(E) \sim |E|^{3/2}$, and the effective dynamic scaling exponent along this critical line is $z_{aw}=6/5$. We find good agreement between the power law scaling of the DOS in the numerics and the perturbative RG expectations along the ThBI to ThDM phase boundary with a quantum critical fan at finite energy anchored by this localization-delocalization transition as shown in Fig.~\ref{fig:phasediagram} (b). 
Since both the exponentially localized Lifshitz states in the ThBI and the power law quasi-localized rare states in the ThDM contribute to a non-zero DOS at $E=0$, our numerical results are consistent with $\rho(0)\neq 0$ at the ThBI to ThDM transition. Moreover, we directly show that the non-analytic behavior in the DOS in the clean limit (at $\mu = \pm 6t$) is rounded out due to non-perturbative effects. Thus, the scaling of the DOS in the quantum critical fan anchored by the ThBI to ThDM QCP only holds at finite energy and inevitably crosses over at low energy to an analytic DOS.
Thus, by tuning disorder strength $W$ or chemical potential $\mu$ we can observe rich crossover behavior at finite energies, which is illustrated in Fig.~\ref{fig:phasediagram}.  

We find that the diffusive phase is well described by a class D nonlinear sigma model only when the disorder strength is larger than the avoided critical strength. As a hallmark of class D diffusive model we observe antilocalization effects on the density of states. The perturbative beta function for the nonlinear sigma model has been known up to four loop orders \cite{Hikami-1981,Wegner1989}. However, the beta function consists of  an alternating series and it is difficult to analytically predict the Anderson localization transition for class D. Thus, our exact numerical calculations are essential for unveiling Anderson localization transition of the class D diffusive metal.

At very large disorder strength we find an Anderson localization transition of the ThDM to an AI, which results from suppressing neighboring hopping with large fluctuations of the onsite potential. Due to anti-localization effects, we find this occurs for a disorder strength much larger than the clean bandwidth as well as the existence of a sharp anti-localization peak in both typical and average DOS in  the ThDM phase. We determine the shape of the mobility edge and the power law governing how the typical DOS goes to zero at the AI transition. Since the ThBI and AI arise from distinct physical mechanisms and are governed by different quantum phase transitions, it is possible that a ThDM phase exists to very large $W/t$ and $\mu/t$ separating the ThBI and the AI [shown schematically as the dashed line in Fig.~\ref{fig:phasediagram} (a)]. 

\section{Model and Method}
\label{sec:model}

We study the following quadratic Hamiltonian of a three dimensional spinless $p_x+ip_y$ superconductor
\begin{align} 
H &= \sum_{\mathbf r} \; \sum_{\hat{\nu}} \;  \left[t \; c_{\mathbf r+\hat{\nu}}^{\dagger} c_{\mathbf r} + i \Delta_{\hat{\nu}} \; c_{\mathbf r+\hat{\nu}}^{\dagger} c_{\mathbf r}^\dagger + \mathrm{h.c.}  \right] 
\nonumber
\\ 
&+  \sum_{\mathbf r} (V(\mathbf r) -\mu) c_{\mathbf r}^{\dagger} c_{\mathbf r},  
  \label{eq:pipHamiltonian}
\end{align}
where $c_{\mathbf r}$ is the fermion annihilation operator at site $\mathbf r $ on a cubic lattice, $\hat{\nu}=\pm\hat{\mathbf x}, \; \pm\hat{\mathbf y}, \; \pm\hat{\mathbf z}$ are coordination vectors for nearest neighbors, $t$ is the nearest neighbor hopping strength, and $\Delta_\nu$ are nearest neighbor pairing amplitudes. For describing $p_x+ip_y$ pairing, we choose $\Delta_{\hat{x}}= \Delta, \; \Delta_{\hat{y}}=-i \Delta, \; \Delta_{\hat{z}}=0$, where $2\Delta$ is the superconducting gap for a clean model. Additionally, $\mu$ and $V(\mathbf r)$ respectively denote the uniform and randomly varying chemical potentials for normal quasiparticles. For our calculations the disorder potential will follow either (i) a Gaussian probability distribution with zero mean and standard deviation $W$ or (ii) a box distribution in the interval $[-W/2,W/2]$ (we will specify which probability distribution we are using in each relevant section). The boundary conditions are taken to be periodic unless otherwise specified with a linear size $L$, and volume $V=L^{3}$. Both disorder distributions give the same qualitative behavior, but differ in some quantitative details.

\subsection{Clean phase diagram}
\label{subsec:clean} 
In the absence of disorder, the mean-field Hamiltonian in the momentum space can be written as $H_0=\frac{1}{2}\; \sum_\mathbf{k} \; \psi^\dagger_\mathbf{k} h(\mathbf{k}) \psi_\mathbf{k}$ where $\psi^\dagger_\mathbf{k}=(c^\dagger_{\mathbf{k}}, c_{-\mathbf{k}})$ is the two component Nambu spinor. The Hamiltonian operator is given by 
\begin{eqnarray}
h(\mathbf{k})=[2t\sum_{j=1}^{3}\cos k_j -\mu ]\tau_3+ \Delta (\sin k_1 \tau_1 +\sin k_2 \tau_2), \nonumber \\
\end{eqnarray} 
where $\tau_j$'s are Pauli matrices operating in particle-hole space, and $h(\mathbf{k})$ satisfies the following particle-hole symmetry condition 
\begin{equation}
\tau_1 \; h^T(-\mathbf{k}) \; \tau_1=-h(\mathbf{k})
\end{equation}
of Altland-Zirnbauer symmetry class D. The quasiparticle spectra for this model are determined by 
$$E_{\pm}(\mathbf{k})=\pm \; [ (2t\sum_{j=1}^{3}\cos k_j -\mu )^2+\Delta^2 (\sin^2 k_1+\sin^2 k_2)]^{1/2}.$$ 

Depending on the ratio $\mu/t$, the clean model can support three gapless and two gapped states, as shown in Fig.~\ref{fig:cleanPhaseDiagram}, and the corresponding anomalous thermal Hall conductivities are displayed in Fig.~\ref{fig:cleanPhaseDiagram}. When $2t< \mu <6t$, the paired state has two Weyl nodes, and the left and right handed Weyl points are respectively located at $\mathbf{k}=(0,0,\pm K_1)$, where $K_1=\arccos \left(\frac{\mu}{2t}-2\right)$. Consequently, the Berry flux through the $xy$ plane will point along the $+\hat{z}$ direction. At low temperatures, the MW fermions display a longitudinal thermal conductivity $\kappa_{xx} \sim T^2$ (arising from residual inelastic scattering effects) and an anomalous thermal Hall conductivity 
\begin{equation}
\kappa_{xy}=\frac{k^2_BT}{6\hbar \pi a} \; \arccos \left(\frac{\mu}{2t}-2\right),
\end{equation}
where $a$ is the lattice spacing. We denote this phase as ThSM2. Notice that $\kappa_{xy}$ vanishes as $\mu \to 6t$ and acquires its maximum value $\frac{k^2_BT}{6\hbar \pi a}$ when $\mu \to 2t$. 

For $\mu >6t$, there is no underlying Fermi surface, and the system is in the BEC regime, where the quasiparticle spectrum is fully gapped, and the paired state acts as a thermal band insulator (both $\kappa_{xx}/T$ and $\kappa_{xy}/T$ vanish in the $T \to 0$ limit). We denote this phase as ThBI. In the vicinity of the QCP between ThSM2 and ThBI located at $\mu=6t$, the low energy excitations are described by
\begin{equation}  
h(\mathbf{k}) \approx -\Delta (k_1 \tau_1 +k_2 \tau_2)+ (6t-\mu - t k^2_3)\tau_3.
\end{equation}
The strongly anisotropic dispersion at the QCP is captured by $E_{AW}(\mathbf{k})=\pm \sqrt{t^2 k^4_3+ \Delta^2_p k^2_\perp}$. At any finite energy (E) we can define two distinct energy dependent correlation lengths: $\xi_3(E) \sim 1/\sqrt{|E|}$ along the $z$ direction and $\xi_\perp \sim 1/|E|$ in the $xy$ plane. Similarly, at finite temperatures we can define two different de Broglie wavelengths or temperature dependent correlation lengths, by replacing $|E|$ with temperature $T$. All the critical properties can be understood in terms of these energy or temperature dependent correlation lengths. As a consequence of such anisotropic scaling, the critical density of states behaves as 
\begin{equation}
\rho(E,\mu=6t,W=0) \sim \xi^{-1}_3(E) \xi^{-2}_\perp(E) |E|^{-1} \sim |E|^{3/2}.
\end{equation}
If we introduce an effective dynamic scaling exponent $z_{\mathrm{aw}}$ such that $\rho(E,\mu=6t,W=0) \sim |E|^{d/z_{\mathrm{aw}}-1}$, we obtain $z_{\mathrm{aw}}=6/5$. The quantum critical fan for this QCP is described by the condition $|E|$ (or $T) > |6t-\mu|$. Outside the critical fan $E$ (or $T) < |6t-\mu|$, the low energy physics of ThSM2 and ThBI is governed by the correlation lengths $\xi_3 \sim 1/\sqrt{|6t-\mu|}$ and $\xi_\perp \sim 1/|6t-\mu|$. Inside the ThSM2 the density of states is reduced and follows the power law $\rho(E) \sim E^2$ reflecting that the dynamic scaling exponent $z=1$ for MW excitations. By contrast, the DOS for the ThBI phase exhibits a sharp spectral gap. We can summarize such critical and off-critical behaviors of the DOS with the following results
\begin{multline}
 \rho(E,\mu^*+\delta\mu,W=0) \\ \sim \begin{cases} E\sqrt{E - \tfrac{\delta\mu}{2}}\Theta(E - \tfrac{\delta\mu}2), & \delta\mu>0, \\
E\left[ \sqrt{E - \tfrac{\delta\mu}2} - \sqrt{\tfrac{|\delta\mu|}2-E}\Theta(\tfrac{|\delta\mu|}2-E) \right], & \delta\mu<0,  \end{cases} 
\end{multline}
where $\Theta(x)$ is the Heaviside step function and $\delta \mu=6t-\mu$. The nodal separation inside the ThSM2 is governed by $\xi^{-1}_3 \sim \sqrt{\delta \mu}$, which also controls how $\kappa_{xy}/T$ vanishes when the QCP is approached from the ThSM2 side. It is important to note that this gives rise to a non-analytic DOS, e.g. the second derivative of the DOS with respect to energy diverges as $\rho''(0) \sim \delta \mu^{-1/2}$.

When we approach the QCP at $\mu=2t$ from the ThSM2 side, the Weyl points move to the Brillouin zone boundaries as $K_1 \to \pi$. Precisely at $\mu=2t$, three flavors of anisotropic critical excitations (similar to the one described above) emerge at $\mathbf{k}=(0,0,\pi)$, $\mathbf{k}=(\pi,0,0)$, and $\mathbf{k}=(0,\pi,0)$. For $-2t<\mu<2t$ we find a new ThSM phase with four MW fermions. The Weyl points are located at $\mathbf{k}=(0,\pi,\pm K_2)$ and $\mathbf{k}=(\pi,0,\pm K_2)$, with $K_2=\arccos \left(\frac{\mu}{2t}\right)$, and we denote this phase as ThSM4. The right and left handed Weyl points are now respectively placed at $k_3=+K_2$ and $k_3=-K_2$. Consequently, the Berry flux due to these four Weyl points through the $xy$ plane is directed along the negative $z$ axis, leading to  
\begin{equation}
\kappa_{xy}=\frac{k^2_BT}{6\hbar \pi a}\left[\pi-2\arccos \left(\frac{\mu}{2t}\right)\right].
\end{equation}
In the range $0<\mu<2t$, $\kappa_{xy}$ decreases from its maximum positive value toward zero. For $\mu<0$, $\kappa_{xy}$ changes its sign and attains the maximum negative value $-\frac{k^2_BT}{6\hbar a}$ at $\mu=-2t$. 

For the QCP located at $\mu=-2t$, the left and right handed Weyl points of ThSM4 merge at $(0,\pi, \pi)$ and $(\pi,0,\pi)$, giving rise to two flavors of anisotropic critical excitations. In addition, a third flavor of critical excitation appears at $(\pi,\pi,0)$. When $-6t<\mu<-2t$, a different ThSM2 phase is realized, where the left and right handed Weyl points are respectively located at $(\pi,\pi,\pm K_3)$, with $K_3=\arccos \left(2+\frac{\mu}{2t}\right)$. The Berry flux through the $xy$ plane due to these Weyl points is directed along the positive $z$ axis and the thermal Hall conductivity for $-6t<\mu<-2t$ becomes
\begin{equation}
\kappa_{xy}=\frac{k^2_BT}{6\hbar \pi a}\left[\arccos \left(\frac{\mu}{2t}+2\right) -\pi\right].
\end{equation}
As we approach $\mu=-6t$, the new set of Weyl points move toward the zone boundaries and $\kappa_{xy}$ gradually goes to zero. For $\mu<-6t$, again there is no underlying Fermi surface and we enter a BEC or ThBI phase. The anisotropic critical excitations at $\mu=-6t$ occur at $\mathbf{k}=\mathbf(\pi,\pi,\pi)$. In  
Appendix~\ref{sec:an},
we present some perturbative analysis of the disordered problem, which will guide our qualitative understanding of the numerically exact results to be presented later in the paper.

\subsection{Typical and average DOS}
In the remainder of the paper, we study the evolution of the clean phase diagram shown in Fig.~\ref{fig:cleanPhaseDiagram} as a function of the strength of disorder.
To study the various regimes in this problem, we compute the average DOS using the KPM technique, which is particularly well-suited here.
For each disorder realization it is defined as
\begin{align}
\rho(E) = \frac{1}{2 V}\sum_i \delta(E -E_i)
\end{align}
where $E_i$ are the eigenstates of the disordered Hamiltonian. 
We then average this quantity over a number of disorder realizations (see Table~\ref{tab:NumericalParameters}).
Additionally, in order to study localization phenomena,  we are also interested in the possibility of a thermal insulating phase (i.e. the AI phase)  driven by very large disorder (i.e.\ a much larger disorder strength than the putative weak-disorder avoided QCPs); this localization of the Majorana BdG quasiparticles can be tracked by computing the typical density of states~\cite{Weisse2006}
\begin{align}
\rho_t(E) = \exp \left[\langle\langle \log \rho_i(E) \rangle\rangle \right]
\end{align}
where $\langle\langle \cdots \rangle \rangle$ denotes a disorder average and the local density of states at site $i$ is defined as 
\begin{align}
\rho_i(E) = \frac1{2V} \sum_n \left|\braket{i| E_n}\right|^2 \delta(E - E_n).
\end{align}
To reach sufficiently large system sizes, we avoid a direct diagonalization of $H$ by using the KPM. 
This method allows us to compute the average and typical density of states for sufficiently large system sizes due to the sparsity of the tight binding Hamiltonian (for technical details see Ref.~\cite{Weisse2006}). 
Essentially, the KPM expands the DOS (and also the local DOS) in terms of Chebyshev polynomials with an appropriate Kernel (here we use the Jackson Kernel). 
The KPM then just needs to use matrix multiplication techniques to find the Chebyshev expansion coefficients.

There are a couple of numerical inputs into this method: We must specify the number of Chebyshev coefficients we are keeping $N_C$, the number of random vectors we use to evaluate the stochastic trace $N_R$ (see Sec.~\ref{sec:modif-stoch-trace} for details about this), the type of disorder distribution used [here we use either Gaussian or box disorder], and how many disorder realizations we take $N_{\rm dis}$.
Unless otherwise stated, we generally use the values indicated in Table~\ref{tab:NumericalParameters} for the average-DOS, and for the typical-DOS we use $L=30$, $N_C=2^{11}$ to $2^{13}$, $N_R=10$ and $N_{\mathrm{dis}}=800$.

\begin{table}[h]
  \centering
    \begin{tabular}{c c c c|c c c c}
      \hline\hline
      \multicolumn{4}{c|}{Box Disorder}& \multicolumn{4}{c}{Gaussian Disorder} \\
      \hline 
      $L$        & $N_C$   & $N_R$  & $N_{\rm dis}$ &  $L$         & $N_C$     & $N_R$ & $N_{\rm dis}$ \\
      \hline
      60         & $2^{11}$ &  9    & 16           &   $\leq 60$  & $2^{11}$  & 9     & $10^4$ \\
      80         & $2^{11}$ &  9    & 8          \\
      $\geq 100$ & $2^{11}$ &  9    & 4         \\ 
     \hline \hline  
    \end{tabular}
  \caption{The numerical input for most calculations.
The size of the system is $L^3$, $N_C$ is the number of Chebyshev coefficients taken, $N_R$ is the number of random vectors in the stochastic trace. 
For box disorder the random vectors are unnormalized while for Gaussian disorder the random vectors are normalized, see Sec.~\ref{sec:modif-stoch-trace}.
Lastly, $N_{\rm dis}$ is the number of disorder realizations.
The system is self-averaging, so not very many are needed.
However, for Gaussian disorder, we use many more in order to see rare-region effects.\label{tab:NumericalParameters}}
\end{table}

\subsection{Modification of stochastic trace}
\label{sec:modif-stoch-trace}
In order to probe the effects of rare regions on the low energy DOS we need to be able to detect an exponentially small DOS on the order of $10^{-5}$. 
However, as shown in \cite{Pixley2016} the KPM method has an artificial background DOS that is flat and sets a lower bound on what size DOS can be accurately detected. 
Here, we show what the origin of this fake KPM background is and establish how to remove it alltogether. 
As a result the lower bound of the DOS is set by the intersection of the broadening of the Dirac delta-functions in the definition of the DOS.

\begin{figure}[!ht]
  \centering
  \includegraphics[width=\columnwidth]{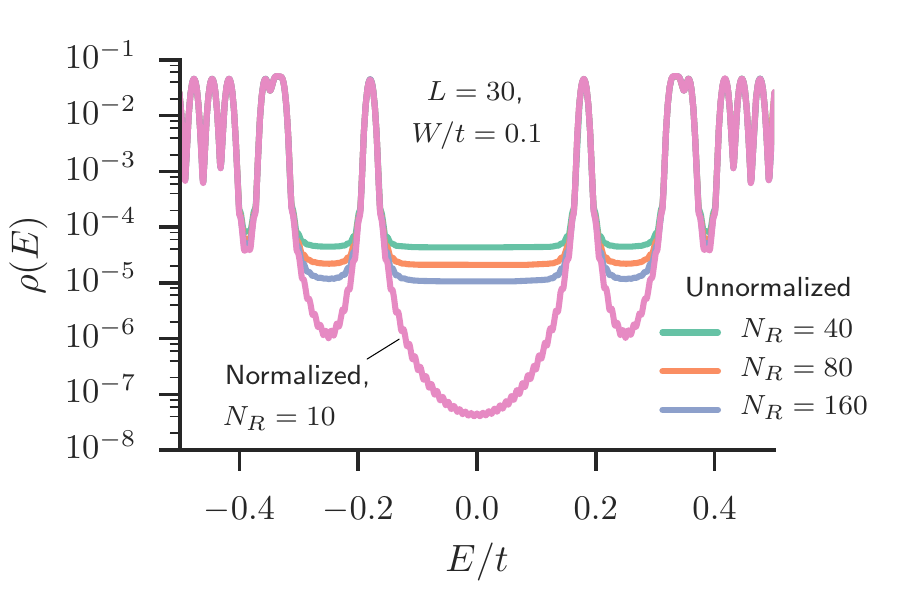} 
  \caption{The stochastic trace evolution of the density of states produces an artificial background as seen by the use of ``Unnormalized'' vectors.
    However, if one just uses a small number of ``Normalized vectors'' much more accurate results at low values of $\rho(E)$ can be found.
    The calculation in this plot was done with 100 disorder realizations and Gaussian disorder at $\mu=0$.
    This effect persists at lower system sizes where it is confirmed by computing the exact trace.}
  \label{fig:KPMbackground}
\end{figure}
Evaluating the KPM requires the trace of a matrix, and the most numerically efficient way to evaluate the trace of a large matrix is with stochastic vectors~\cite{Weisse2006}.
Usually, one takes Gaussian sampled vectors $\ket{r} = \sum_i \xi_{ri}\ket{i}$ such that
\begin{align}
 \braket{\xi_{ri}^*\xi_{r'i'}} = \delta_{rr'}\delta_{ii'}
\end{align}
where $r$ denotes the random vector and $i$ its component.
However, this means that the vectors we are sampling over are only normalized \emph{on average}, and the fluctuations away from average are what lead to the artificial KPM background that masks the real value of the DOS, as shown in Fig.~\ref{fig:KPMbackground}.
As has been pointed out~\cite{Hams2000,Iitaka2004}, normalization of the stochastic vectors used in the evaluation of the trace can reduce the statistical noise in the data.
To implement this formally, we insist that each individual vector is normalized. 
Thus after sampling the vectors, we impose the normalization condition
\begin{align}
 \sum_i|\xi_{ri}|^2 = 2V.
\end{align}
We demonstrate how this works in practice in Fig.~\ref{fig:KPMbackground}: for $\rho(E) \gtrsim 10^{-4}$ all vectors work (normalized or unnormalized), but the unnormalized vectors hit the artificial KPM background between $10^{-4}$ to $10^{-5}$ that is just marginally improved by taking more random vectors. 
On the other hand, the normalized vectors work down to very low numbers dictated by, in this case, the tail of the Gaussian-broadened states (due to the Jackson Kernel). 
Even if we use many more normalized random vectors, the results shown in Fig.~\ref{fig:KPMbackground} do not change substantially.
Furthermore, at smaller sizes where an exact trace is computationally reasonable, the agreement is much better.
This technical advancement in our implementation of KPM is substantial as it now opens the door for using the KPM to detect the existence of low energy rare region effects, which prior to our work, was only possible at moderate disorder strengths above the fake KPM background. Our modification essentially gets rid of the constraint arising from the KPM-induced artificial DOS background problem.

\section{Non-perturbative effects at weak disorder}
\label{sec:non-pert}

Since the DOS for both the MW fermions and the critical excitations at the anisotropic QCP, vanish faster then $|E|$, weak disorder is expected to be an irrelevant perturbation (in the RG sense) to the clean phase diagram in Fig.~\ref{fig:cleanPhaseDiagram}. The self-consistent Born approximation (SCBA) calculation also suggests that the ballistic MW fermions remain stable (i.e., $\rho(E=0)=0$) up to a critical strength of disorder $W_c(\mu)$. For $W>W_c(\mu)$, disorder induces a finite density of states at zero energy, giving rise to a diffusive thermal Hall metal (i.e. the ThDM phase). The perturbative irrelevance of disorder at the anisotropic QCP (i.e. $\mu=\pm 6t$) also suggests the existence of a perturbatively accessible disorder driven (metal to insulator) transition $W_I(\mu)$, consistent with the SCBA. However, this picture is drastically modified by non-perturbative effects of disorder as we show in this section. We find numerically that the DOS is \emph{analytic} across the entire $W-\mu$ phase diagram and our results are consistent with the DOS always being non-zero in the thermodynamic limit.
Despite its shortcomings, the perturbative RG does correctly capture the shape of the phase boundary separating the ThDM and ThBI phases and provides a quantitative description of the power law scaling of the DOS and the average band gap. Due to its technical nature and to avoid confusion with full numerical solution (that incorporates all effects) we present the perturbative RG in Appendix~\ref{sec:an}.

\subsection{Thermal band insulator $|\mu|>\mu_I(W)$}

\begin{figure}[!ht]
  \centering
 \includegraphics[width=\columnwidth]{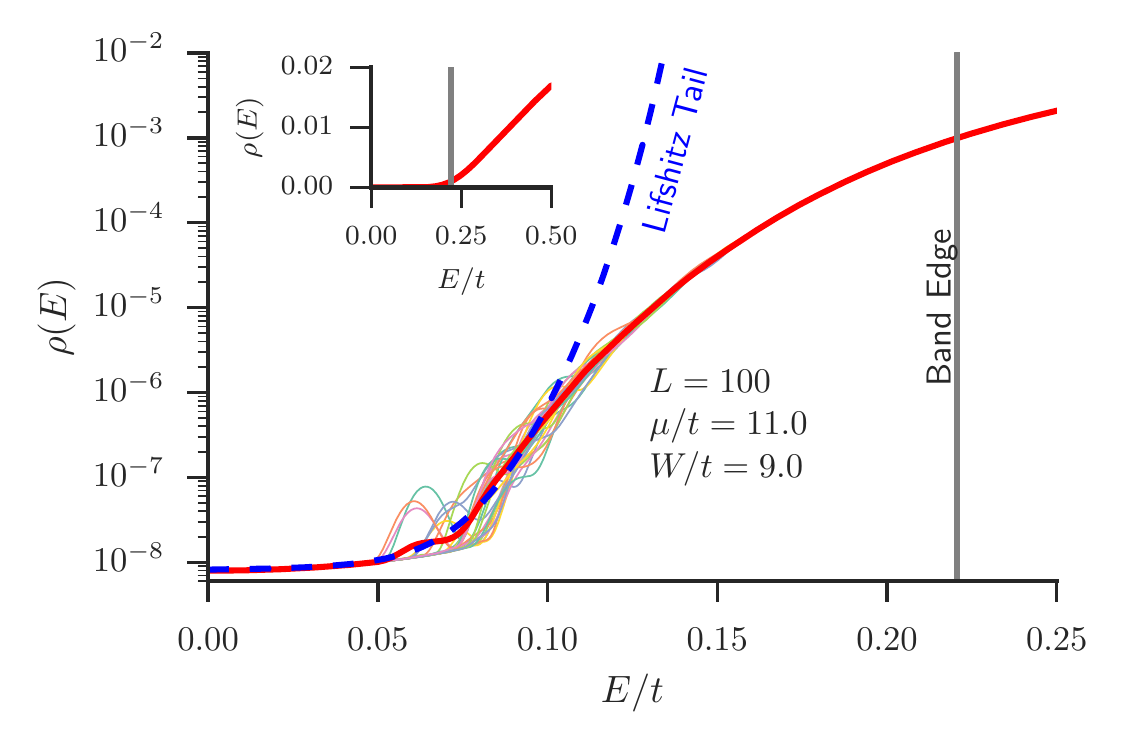} 
  \caption{Deep within the insulating phase, we fit the Lifshitz Tail due to rare region effects: $\rho(E) - \rho(0) \approx A e^{b|E - E_0|^{-3/2}}$ for box disorder (with $A\approx 5.31\times 10^{-12}$, $b\approx 0.209$, and $E_0\approx 0.233 t$). The thin individual lines are averages over $1000$ realizations while the thick (red) line is an average over 28,000 realizations---all calculations are done with $N_c=2^{13}$. The thin lines clearly show rare contributions to the DOS deep in the average band gap.
  $E_0$ is a good approximation for the average band edge, drawn as a vertical line in both plot and inset. (Inset) Same data on a linear-linear scale. 
  \label{fig:LifshitzTail}}
\end{figure}

We begin by discussing the low energy eigenstates inside the band gap of the thermal band insulator in the phase diagram of Fig.~\ref{fig:phasediagram}. 
For weak disorder, mid-gap states fall randomly inside the band gap that induce Lifshitz states, which are fairly well understood~\cite{Mieghem1992,Kramer1993}. 
These eigenstates are \emph{exponentially localized} around the sites (or cluster of sites) with a very large disorder strength. 
These eigenstates round out the gap in the DOS and give rise to an exponentially decaying energy dependent DOS that goes like $\rho(E)\sim A(E) e^{-B(E)}$ \cite{Yaida2016} (e.g.\ $\rho(E)\sim A e^{-B |E-E_0|^{-3/2}}$ for box disorder and $\rho(E)\sim A e^{-B|E-E_0|^{1/2}}$ for Gaussian disorder \cite{Mieghem1992,Kramer1993} in three dimensions).
We demonstrate this in the ThBI phase in Fig.~\ref{fig:LifshitzTail} where we fit the DOS to the Lifshitz form near the band edge.
In the ThBI phase the zero energy eigenstates are Anderson localized insulating states and their contribution to the zero energy DOS is negligible, but strictly speaking non-zero.  Thus, despite the inability to probe such zero energy states numerically, our data satisfying the Lifshitz form is consistent with $\rho(0) >0$ in the disordered ThBI phase.
For sufficiently large disorder strength, these mid-gap states become sufficiently dense to completely fill in the average bulk band gap. In this limit these Lifshitz states at $E=0$ develop sufficient overlap eventually driving a quantum phase transition from the ThBI to a ThDM. 
We explore this transition non-perturbatively in Section~\ref{subsec:thdmtobi} and its scaling properties in Section~\ref{sec:thsm-thbi}.

\subsection{Thermal semi-metal regime $|\mu|<\mu_c(W)$}
In this section, we provide numerical evidence at weak disorder that non-perturbative rare-regions exist in the current model (despite the strict particle hole symmetry) and contribute a non-zero DOS at $E=0$. 
We focus on a gaussian distribution of disorder because the unbounded tails of the distribution lead to large local fluctuations of the potential enhancing the probability of generating a rare event~\cite{Nandkishore2014, Pixley2016a}.
Moreover, as these events are rare we use $10,000$ disorder realizations to find a statistically significant result.

\begin{figure}
  \centering
 \includegraphics[width=0.45\textwidth]{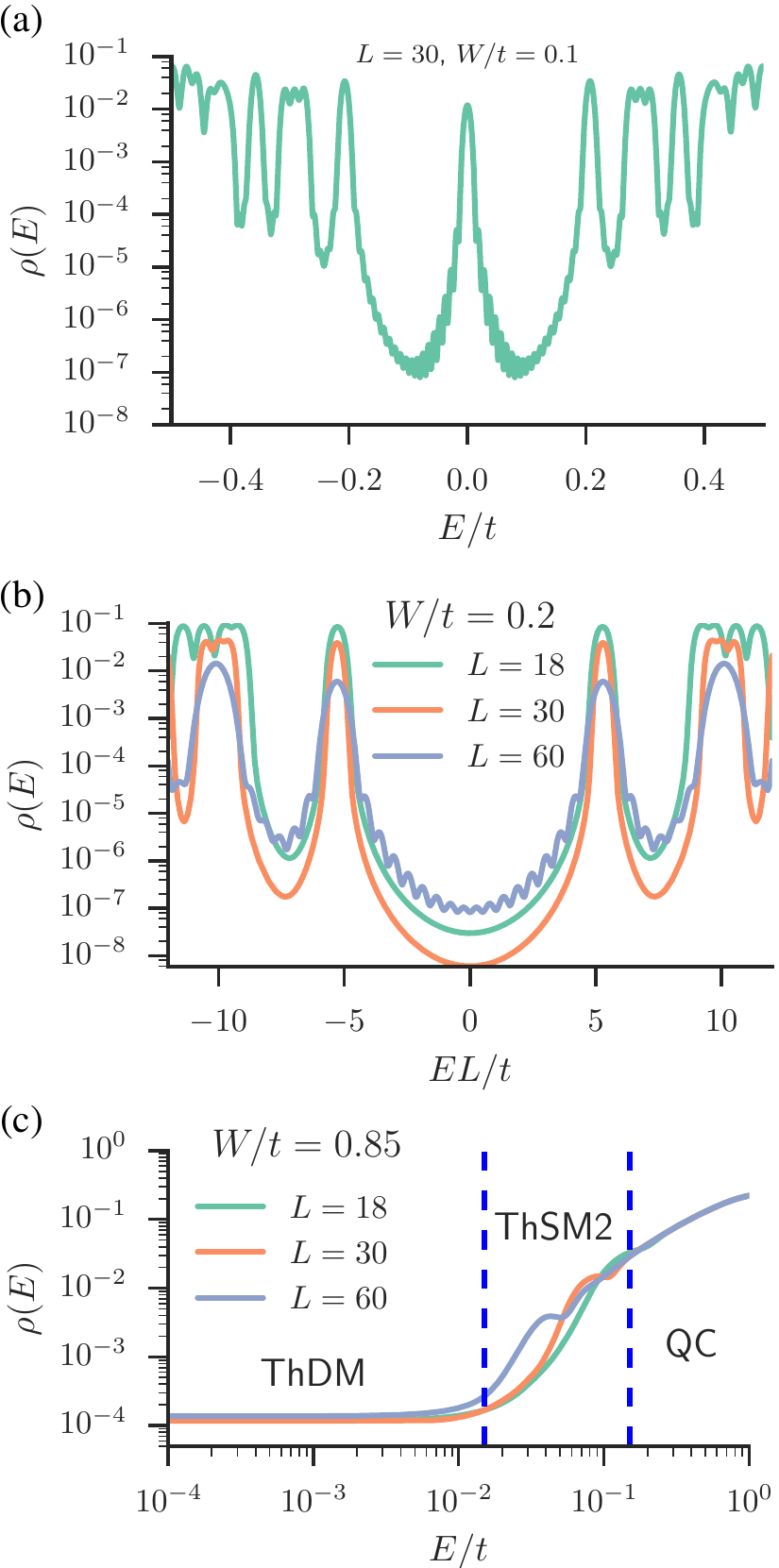}
  \caption{The finite size DOS at weak disorder. (a) Depending on the physics we are looking at, it is useful to sometimes keep one Weyl peak at zero energy as shown here. (b) The DOS for various $L$ with Gaussian disorder as a function of $LE$.  At weak disorder the peaks are spaced like $1/L$ and therefore are composed of perturbatively dressed Weyl states that are smoothly connected to their $W=0$ counterparts.
    There are no states at zero energy, the finite value of the DOS at $E=0$ is solely due to the overlap of the two Weyl peaks at $EL/t \approx 5$ broadened by the KPM.  (c) The DOS versus $E$ computed with a gaussian disorder distribution displaying the cross over regimes at low $E$. At very low energies the density of states is non-zero and essentially $E$-independent displaying the ThDM regime. At finite $E$, the ThSM regime is intact with characteristic $\rho(E)\sim E^2$ dispersion, and at large $E$, we start to see the critical fan $\rho(E)\sim |E|$.}
  \label{fig:DiracStatesvsEL}
\end{figure}
As shown in Ref.~\cite{Pixley2016}, in order to diagnose rare region effects and eliminate the finite size effects on the DOS in the SM regime, we need to move the zero energy states away from $E=0$ so that for $W=0$, $\rho(0)=0$ at all $L$. 
In short, by modifying boundary conditions alone we can move from a situation with a state at $E=0$ as shown in Fig.~\ref{fig:DiracStatesvsEL} (a) and move the lowest energy Weyl state maximally from zero as shown in Fig.~\ref{fig:DiracStatesvsEL} (b).
To accomplish this in Nambu space we either consider $L$ that is not a multiple of four with periodic boundary conditions or we introduce antiperiodic boundary conditions, which both move the clean (i.e.\ disorder-free) states away from zero energy.
In particular, for anti-periodic boundary conditions $\psi(0) = - \psi(L \mathbf e_j)$ for the directions $j=x,y,z$ when $L$ is a multiple of four~
\footnote{
  The location of disorder-free states in $k$-space depends heavily on the size of the system due to the location of the nodes going as $k_z^* = (2n+1)\pi/2 = 2\pi m/L$ for integers $n$ and $m$. 
  Therefore, we have zero-energy states appear as nodes when  $L$ is a multiple of 4 and do not appear when $L$ is even but not a multiple of 4. 
  Therefore, in the former case we use antiperiodic boundary conditions while in the latter we use periodic boundary conditions in the $z$-direction only.}.
In the disorder-free limit with anti-periodic boundary conditions 
the eigenstates labeled by ${\bf k}$ are shifted to ${\bf k}\rightarrow {\bf k} + \pi/L(1,1,1)$. 
After performing this shift, the lowest lying Weyl state is maximally moved away from zero energy.
Also, we want to minimize the broadening of each energy eigenvalue by disorder, and we do this by enforcing $\sum_r V(r) = 0$ for each disorder realization.
Here, we are seeking an exponentially small $\rho(0)$ and therefore the modified stochastic trace (see Sec.~\ref{sec:modif-stoch-trace}) is essential to be able to access a DOS of such a small magnitude.

\begin{figure}[h!]
  \includegraphics[width=0.45\textwidth]{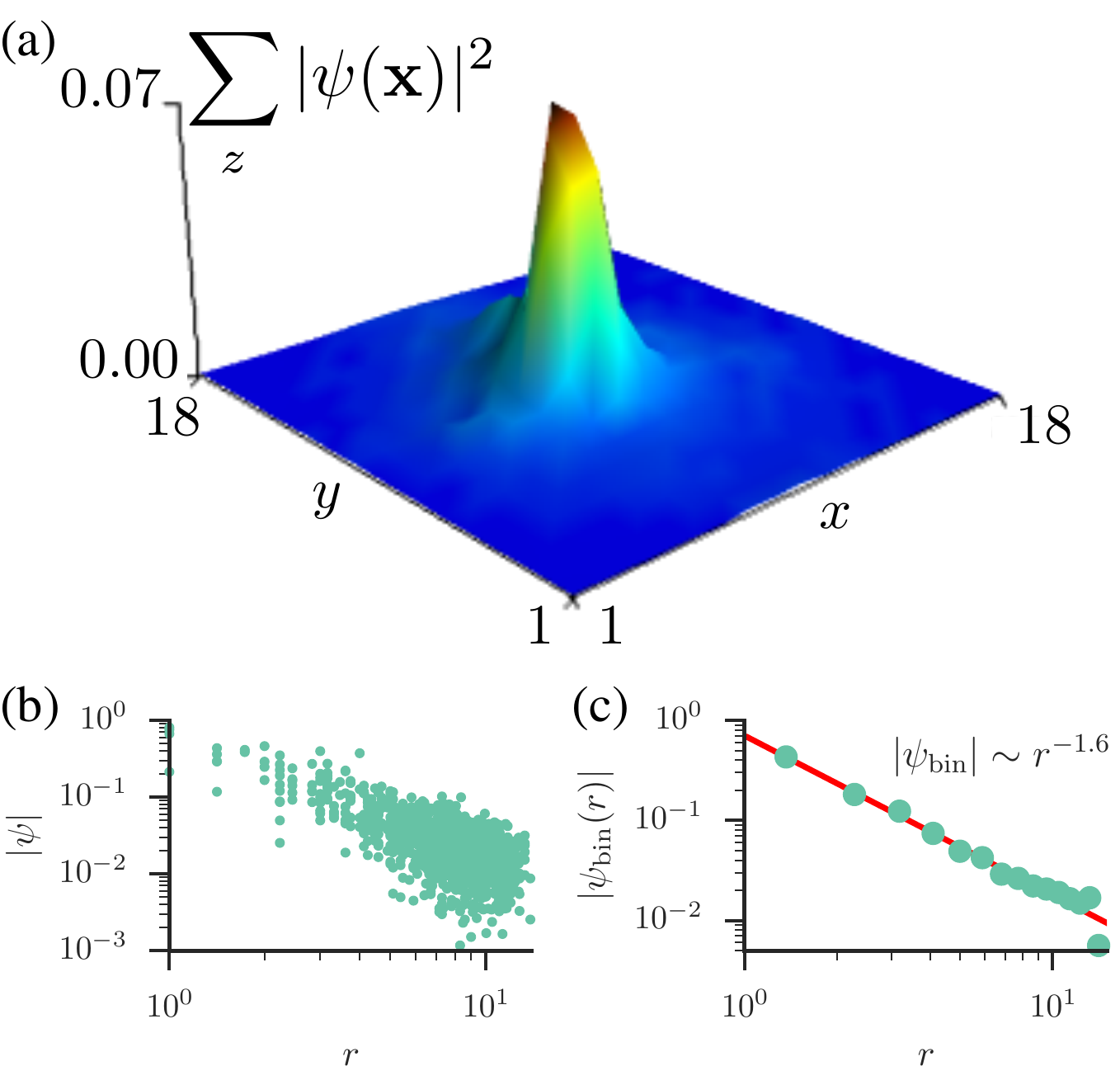}
  \caption{(a) Projected probability density $\sum_z|\psi(x,y,z)|^2$ in the $xy$-plane of a rare wavefunction that corresponds to a state in the low energy tail of the DOS with $W/t=0.8$, $L=18$, and periodic boundary conditions.
   (b) A scatter plot of the wave function decay away from its maximum value on a log-log plot, showing a clear power-law trend.
 (c) The scatter plot data is binned and is fit to $a/r^x$ showing a clear power law decay at small $r$ with $x=1.6$ and the rare wavefunction is indeed quasi-localized.}
\label{fig:localizedstate}
\end{figure}

For very weak disorder the low energy DOS is well described by Weyl peaks that are broadened and move in energy due to disorder, see Figs.~\ref{fig:DiracStatesvsEL} (a) and (b). 
At slightly larger disorder, (where we can find a statistically significant amount of rare states) the low energy DOS is an essentially flat background (between the Weyl peaks) that extends to $E=0$, see Fig.~\ref{fig:DiracStatesvsEL} (c).
These Weyl peaks are well described by perturbation theory in disorder and are essentially perturbatively renormalized Weyl states~\cite{Pixley2016}, which are spaced like $~1/L$ (see Fig.~\ref{fig:DiracStatesvsEL}).
The flat background DOS on the other hand originate from quasi-localized rare eigenstates.
Focusing on a disorder sample that produces a low energy state contributing to the flat part of the DOS, we compute the two component spinor wave function  $\psi(x,y,z)$ of the this state using Lanczos on $H^2$ for $\mu=0$, $L=18$ and $W=0.8t$.
Note that for $L=18$ and periodic boundary conditions this places the first low energy state (in the clean limit) to be at $E_0\approx 0.3t$ and thus satisfies $\rho(0)=0$ for this value of $L$.
We project the probability amplitude into two dimensions via $\sum_z|\psi(x,y,z)|^2$ for plotting purposes.
We find that the wavefunction is quasi-localized in real space [see Fig.~\ref{fig:localizedstate}(a)] about two sites one with a value of $V_i\sim4W$ and the other $V_i\sim2W$, with a probability $\sim \exp(-10)$ which is indeed a rare eigenstate relative to $V_i \sim W$. 

We now compute the decay of the wavefunction from its maximal value.
To do this we first compute the distance to the site where the wave function has its maximum $\br_{\mathrm{max}}$ and then compute the distance from it $\psi(r)\equiv \psi(|{\bf x} - \br_{\mathrm{max}}|)$ (respecting the periodic boundary conditions $|x^{\mu}- r_{\mathrm{max}}^{\mu}| <L/2$), the scatter plot of this is shown in Fig.~\ref{fig:localizedstate} (b).
We then bin the wavefunction along $r$ and compute its power law decay, which leads to one of our main results, namely for this particular rare state we find
\begin{align}
\psi(r) \sim \frac{1}{r^{1.6}}.
\end{align}
This power law decay can vary from one disorder sample to the next but we do find good agreement with the expected analytic prediction of $1/r^2$ (Ref.~\cite{Nandkishore2014}). 
Thus, we conclude that these quasi-localized rare eigenstates are unaffected by the the presence of particle-hole symmetry.

It is important to contrast these quasi-localized eigenstates in the ThSM regime with the exponentially localized Lifshitz states in the Anderson localized ThBI phase.
These quasi-localized eigenstates have level repulsion \cite{Pixley2016} and are not Anderson localized.
Therefore focusing on weak disorder and tuning $\mu$ across $\mu_I(W)$ (i.e. the ThDM to THBI transition) is a true metal to insulator quantum phase transition where the power law quasi-localized rare states are converted into exponentially localized Anderson insulating zero energy states.

\begin{figure}[!ht]
\includegraphics[width=\columnwidth]{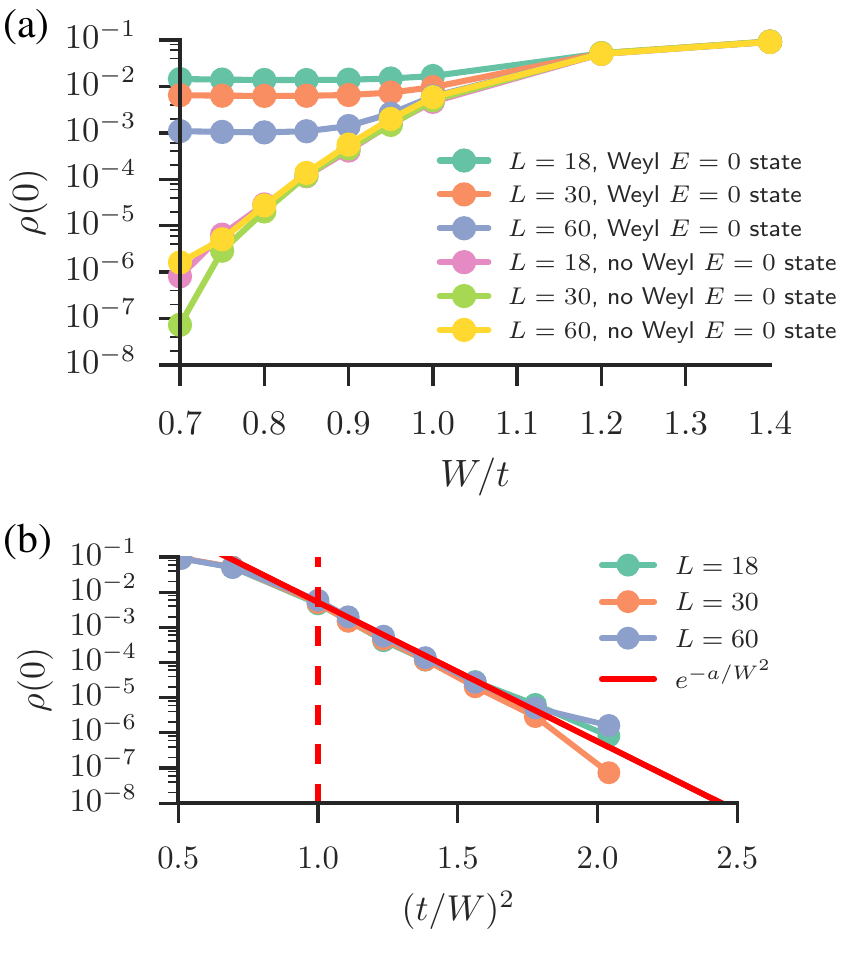} 
\caption{(a) The DOS at zero energy $\rho(0)$ as a function of disorder strength $W$. A Weyl state at zero energy produces a very large finite size effect in $\rho(0)$, whereas the DOS without a Weyl state becomes $L$ independent at weak disorder ($W\gtrsim0.75t$). With a Weyl state at zero energy, we can pinpoint the crossover due to the avoided transition to occur roughly around $W/t\approx 1.0$ (for Gaussian disorder by finite size scaling). 
  (b)  Without a Weyl state at $E=0$, rare states begin populating the low-$|E|$ DOS and they appear as an $L$ independent DOS. The data is well fit to the rare region form $\rho(0)\sim \exp(-a/W^2)$ and our results are consistent with only the ThDM phase persisting at zero energy for weak disorder.
  Even without an $E=0$ state, we hit an artificial background DOS due to the Gaussian broadening of the Dirac-delta function in the DOS of nearby states around $W/t\approx 0.7$.}
\label{fig:rho0RareStates}
\end{figure}

Having identified the eigenstates that make up the low energy relatively flat $L$ independent background DOS extending to $E=0$ we are in a good position to determine the evolution of $\rho(0)$ versus W as seen in Fig.~\ref{fig:rho0RareStates}.
In Fig.~\ref{fig:rho0RareStates} (b) we plot the $L$ independent background DOS with an excellent fit to the rare region form~\cite{Nandkishore2014}
\begin{equation}
\log \rho(0) \sim {(t/W)}^{2}.
\end{equation}
The data is well fit to this form \emph{over 4 orders of magnitude} of $\rho(0)$ ranging from $W=0.75t$ to $1.2t$.
For disorder strength less then $W\approx 0.7$ the rare states are generated with such a low probability that we cannot accurately estimate their contribution to $\rho(0)$ on these size samples for this number of disorder realizations.
The non-zero DOS at $E=0$ has converted the ThSM into a ThDM at $E=0$. In Fig.~\ref{fig:rho0RareStates} (a) we compare the data with anti-periodic boundary conditions to the case of using periodic boundary conditions, which for this $L$ give rise to a Weyl peak centered at $E=0$, inducing a large finite size effect and obscuring the DOS at weak disorder.
The data with periodic boundary conditions does help however in providing an estimate of the avoided QCP.

\begin{figure}
  \includegraphics[width=\columnwidth]{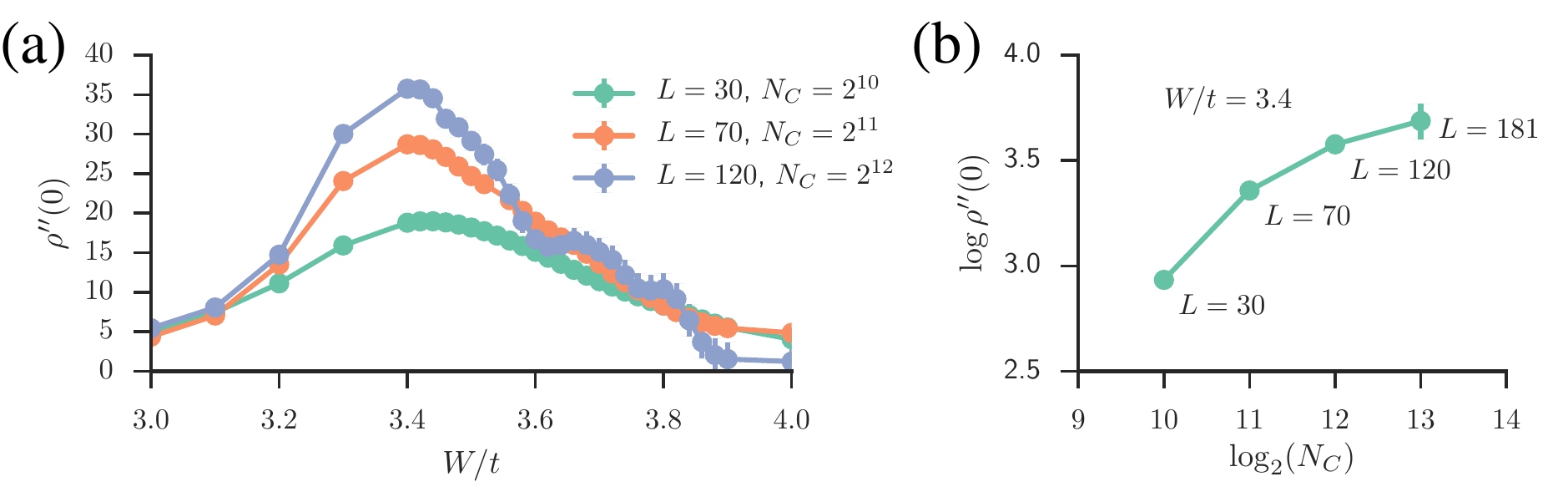}
  \caption{(a) The second derivative of the DOS at zero energy at $\mu=0$ and as a function of disorder strength $W/t$ for \emph{box} disorder. The peak describes the avoided QCP and remains finite as we we saturate the value of $L$ at each particular $N_C$. (b) Focusing at $W/t=3.4$, and increasing with $N_C$, we eventually see the value of the peak saturating suggesting the critical point is indeed avoided. This data has been averaged over the twisted boundary conditions. \label{fig:logd2rhoPeakThDM}}
\end{figure}

It is natural to expect that the non-zero $\rho(0)$ rounds out the QCP into an avoided transition. To show this explicitly we study the strength of the avoidance by assuming that the DOS is always analytic. This implies that
\begin{equation}
\rho(E) = \rho(0) + \frac{1}{2!}\rho''(0)E^2 + \frac{1}{4!}\rho^{(4)}(0)E^4  +\dots.,
\label{eqn:aqcp}
\end{equation}
and if the DOS becomes non-analytic $\rho''(0)$ and $\rho^{(4)}(0) \rightarrow \infty$. Therefore, we use the size of the second derivative (with respect to energy) of the zero energy DOS $\rho''(0)$ to measure the strength of avoidance~\cite{Pixley2016a}. We compute $\rho''(0)$ directly using the KPM~\cite{Pixley2016a}. As shown in Fig.~\ref{fig:logd2rhoPeakThDM} we find that for box disorder $\rho''(0)$ is saturated in both system size and KPM expansion order. We conclude that the DOS is analytic and the ThSM to ThDM QCP is avoided.

Similar to what has been discussed in Refs.~\cite{Pixley2016,Pixley2016a}, these rare states lead to a destruction of the ThSM at low energy, but as we will explore later there is still a regime at finite energy where the effective semimetallic scaling in the DOS can be seen.
We sketch this for the QCPs and avoided QCPs relevant in this work in Fig.~\ref{fig:phasediagram}(b).
Rare regions turn the critical point to an avoided critical point, but we can still probe higher energy cross over features of the DOS that are dictated by the hidden QCP. Thus, although strictly from a theoretical viewpoint any disorder destroys the ThSM phase creating the ThDM phase, for all practical purposes an effective ThSM regime can still be observed in the crossover behavior at higher energy and lower disorder.

\subsection{Thermal diffusive metal to band insulator transition $\mu_I(W)$}
\label{subsec:thdmtobi}
We now focus on the evolution of the clean QCP separating the ThSM2 and ThBI in the presence of disorder. As we have discussed in Section~\ref{subsec:clean}, the anisotropic QCP at $\mu_I(W=0) = \pm 6 t$ has a non-analytic DOS $\rho(E) \sim |E|^{3/2}$, which gives $\rho''(0) \sim | \mu-\mu_I(0)|^{-1/2}$. We now study the evolution of this point in the presence of disorder. In Appendix~\ref{sec:an} we treat the effect of disorder on the anisotropic QCP within a perturbative RG approach. We find that (perturbatively) the QCP survives in the presence of disorder with a renormalized phase boundary, and the non-analytic behavior in the DOS remains with the same critical exponents as in the clean limit. As we show in Sec.~\ref{sec:thsm-thbi}, the RG predictions provide an accurate estimate of the power law scaling in both the DOS and the average band gap. Here, however, we are concerned with the asymptotic low energy behavior of the DOS and whether or not the non-analytic behavior that the power law implies holds all the way down to $E=0$. 

To address this numerically, we focus on weak disorder and vary the chemical potential passing from the ThDM to ThBI phase. This transition is an Anderson localization transition and will be characterized by non-self averaging quantities (such as the typical DOS) developing single parameter scaling. However, we are concerned with following the clean  quantum critical properties in the presence of disorder and therefore focus on the average DOS. Near the ThDM to ThBI transition, it is natural to expect that there will be some non-trivial interplay between rare regions that are either exponentially localized Lifshitz states or quasi-localized power law states. Since both of these effects are inherently non-perturbative, we study the strength of the non-analyticity in the DOS at weak disorder by computing $\rho''(0)$ within the KPM.

\begin{figure}
  \includegraphics[width=\columnwidth]{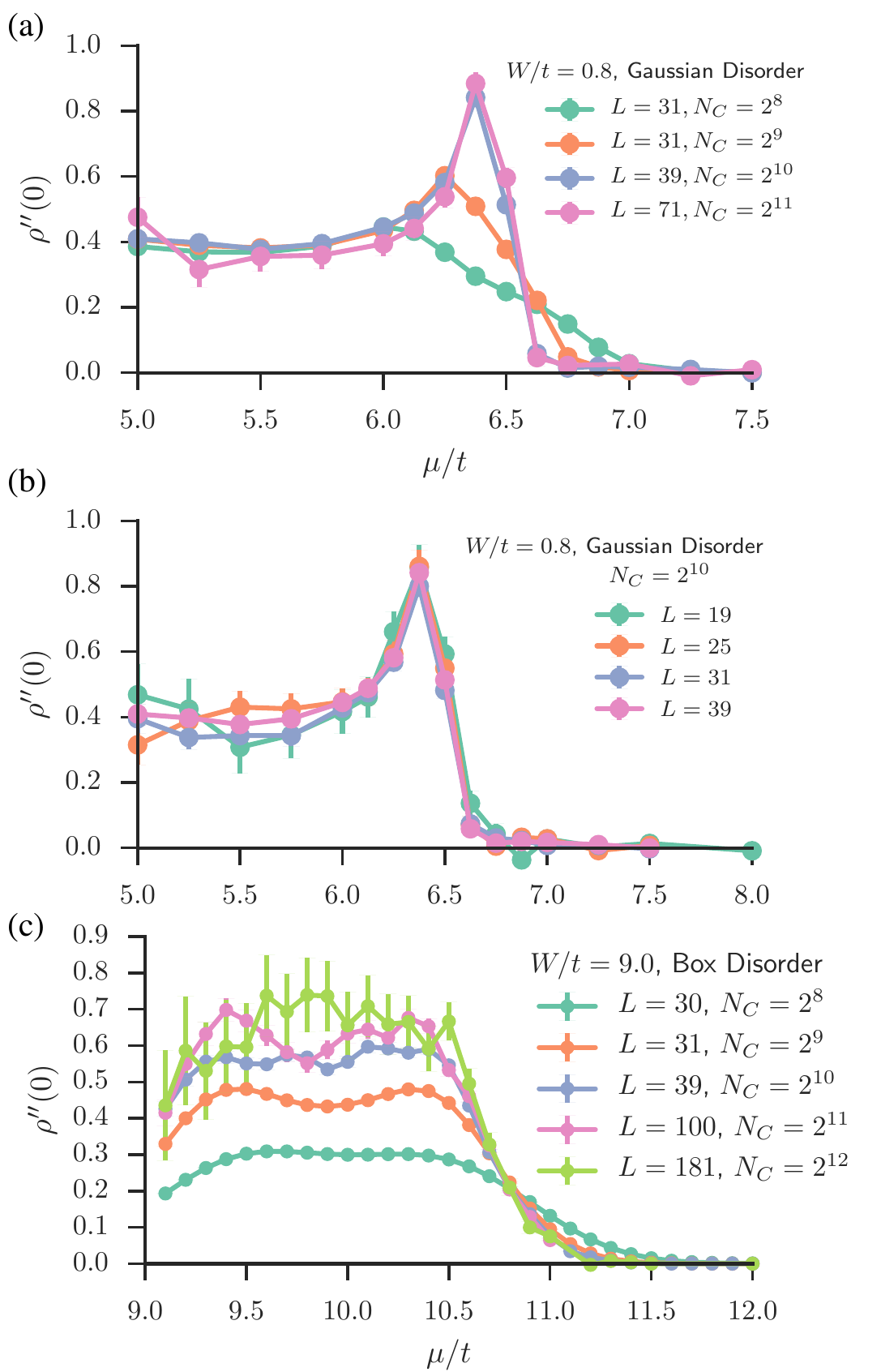}
  \caption{(a) The second derivative of the DOS across the transition from ThSM to ThBI. At $W=0$, this peak is non-analytic, but for finite disorder, we find that it saturates as we increase $N_C$. We see this saturating clearly in (b) where the peak's height is independent of the size for $N_C=2^{10}$.
  (c) The second derivative of the DOS as a function of $\mu/t$ across the transition from ThDM to ThBI. We saturate the second derivative of the DOS across the transition. The results are suggestive that the DOS remains analytic across the transition, but are inconclusive as to whether an intermediate ThSM regime remains between the ThBI phase and ThDM regime.
  Each of these was computed with 1,000 disorder realizations.\label{fig:SaturateSMtoBI}}
\end{figure}

As shown in Figs.~\ref{fig:SaturateSMtoBI} (a) and (b), we find that there is a sharp peak in $\rho''(0)$ which provides an accurate estimate of $\mu_I(W)$.  We find that the peak is saturated in both $L$ and $N_C$. Thus, by removing all of the extrinsic rounding due to finite size effects, we conclude that non-perturbative effects of disorder give rise to an intrinsic rounding that suppresses the divergence of $\rho''(0)$ and the DOS remains analytic at the ThDM to ThBI transition. There is a regime at moderate disorder strengths, where the avoided quantum critical line $\mu_c(W)$ approaches the phase boundary $\mu_I(W)$. As shown in Fig.~\ref{fig:SaturateSMtoBI}(a) and (b), in this regime our data does reveal the existence of two peaks, suggestive that the two lines never intersect.  For very large $W$, we find these peaks become very broad and are smeared out, Fig.~\ref{fig:SaturateSMtoBI}(c). 

\section{Phase Diagram}
\label{sec:phase}

\begin{figure}
  \includegraphics[width=\columnwidth]{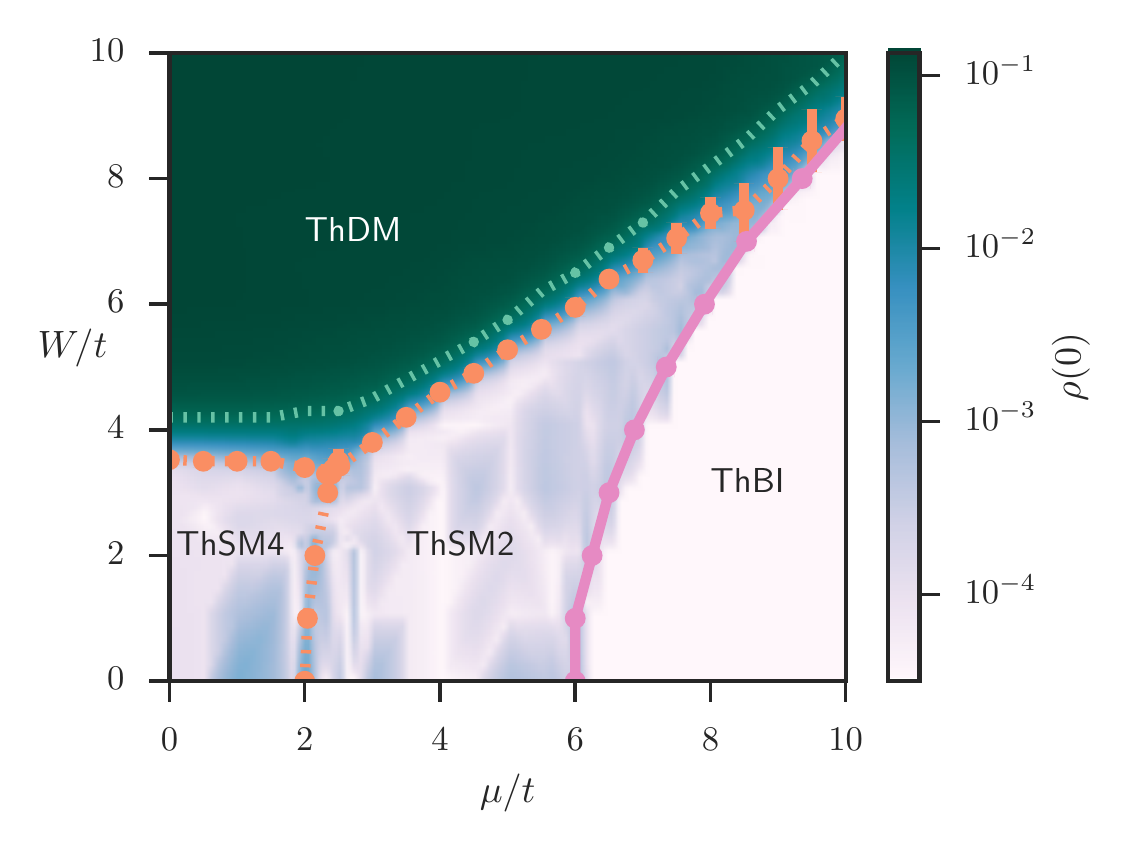}
  \caption{Phase diagram for axial chemical potential disorder in this system with $\Delta = t$. The respective phases are ``metallic'' or ``insulating'' in terms of the Bogoliubov quasiparticles (leading to thermal insulators or thermal conductors), but thermally metallic phases host different regimes that can appear semimetallic at non-zero energy or diffusive at low energies with avoided quantum critical lines that ``separate'' the two as we plot here in dashed orange.
    ThSM$n$ represents a thermal semi-metallic regime with $n$-nodes. 
    ThBI is the thermal band insulating phase while ThDM is the thermal diffusive metal regime.
    The dashed line in the ThDM phase is the onset of a zero-energy anti-localization peak in the DOS associated with the this class D system~\cite{Senthil1998,*Senthil1999,*Senthil1999a}. 
    The color plot is the value of the order parameter $\rho(0)$ and is interpolated from all of the data computed in this paper. 
    The noisy nature of $\rho(0)$ in the semimetallic phases is due to finite size effects and sparsity of data. 
    The color scale uses a lower bound of $\rho(0)\geq 3\times10^{-4}$.\label{fig:basicphasediagram}}
\end{figure}

\begin{figure}
  \includegraphics[width=0.4\textwidth]{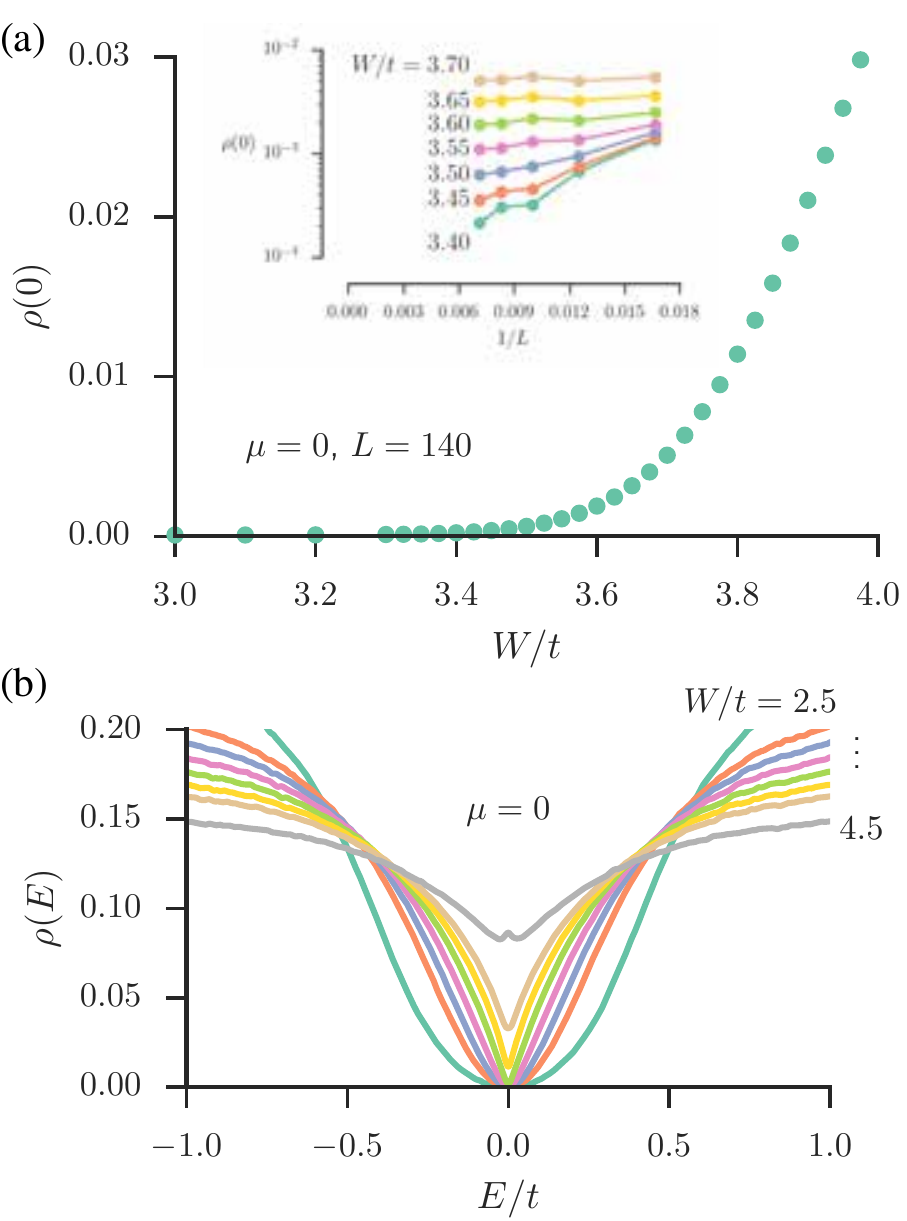}
  \caption{The crossover from ThSM4 to ThDM at $\mu=0$ for box disorder. (a) The rise of $\rho(0)$ as a function of disorder on a linear-linear scale.
  In the inset, we see the saturation of $\rho(0)$ in terms in the ThDM regime for large $L$.
 (b) The DOS changes across this transition, going through a quantum critical region where the DOS is linear near $E=0$.
  Notice that a peak begins to develop for the higher disorder.
  We investigate this peak in Appendix~\ref{sec:ThermalDiffMetal}.
  \label{fig:mu0SMtoDM}} 
\end{figure}

Now that we have determined the physics of this model at weak disorder we now move onto establishing the full thermal phase diagram as a function of the chemical potential ($\mu$) and disorder ($W$) in Fig.~\ref{fig:basicphasediagram}. 
We will then move onto study the avoided criticality between ThSM and ThDM regimes as well as the quantum phase transitions separating the zero energy ThDM and ThBI phases in detail in Sections~\ref{sec:qpt} and~\ref{sec:thsm-thbi} respectively.
In order to determine the regimes and phases in Fig.~\ref{fig:basicphasediagram} using the density of states alone, we use slightly different techniques to find the various transitions and crossovers.
Generally, we try to keep the Weyl state near zero-energy as in Fig.~\ref{fig:DiracStatesvsEL}(a) to study the avoided critical phenomena at finite energy in the quantum critical fan [as illustrated in Fig.~\ref{fig:phasediagram}(b)], and we use a box potential with $V(\mathbf r)\in [-W/2,W/2]$.

We find that each ThSM regime is only present at non-zero energy, which then crosses over to a diffusive metal upon lowering $E$ or increasing $W$. 
For fixed $E$, tuning $W$ allows us to pass through a quantum critical crossover regime that is anchored by the avoided QCP line $W_c(\mu)$, (see the thin solid orange line in Fig.~\ref{fig:phasediagram} and the dotted orange line in Fig.~\ref{fig:basicphasediagram}).
In the ThSM regime, we find the DOS has the form $\rho(E) \sim E^2$ for $E>E^*$ [where $E^*$ is the crossover energy to the ThDM regime with an $L$-independent zero energy DOS that is roughly $E$ independent, see Fig.~\ref{fig:mu0SMtoDM}(a) and (b)], with $\rho(0)$ decreasing with increasing $L$ due to the Weyl peak centered about $E=0$. Near the avoided QCP the low energy DOS goes like $\rho(E) \sim |E|$ for $E>E^*$ [see Fig.~\ref{fig:mu0SMtoDM} (b)], which defines the quantum critical regime.
At finite energy, upon crossing the avoided critical line $W_c(\mu)$ the model 
crosses over
into the ThDM with a finite DOS at zero energy that becomes $L$-independent.  
Even though this shares similarities with its non-superconducting counterparts \cite{Fradkin1986,Fradkin1986a,Shindou2009,Goswami2011,Ominato2014,Bitan-2014,Nandkishore2014,Leo-2015,Syzranov2015,Altland-2015,Syzranov2016,Kobayashi2014,Sbierski2014,Pixley2015-prl,Sbierski2015,Pixley2015,LiuShindou,Bera2016,Shapourian2015,Pixley2016,Pixley2016a,Louvet-2016}, the superconducting case is quite different due to particle-hole symmetry. 
For example, unlike the metallic case, the beta function from the non-linear sigma model analysis does not possess a zero~\cite{Senthil1998,*Senthil1999,*Senthil1999a}, which is suggestive that disorder cannot easily localize the zero energy state. 
Within our calculations this gives rise to a peak in the low energy average DOS that goes like $-\sqrt{E}$ at sufficiently low energy deep in the ThDM regime, as described in detail in Appendix~\ref{sec:ThermalDiffMetal}. 
We find that this peak does not occur until crossing a boundary that is always slightly larger then $W_c(\mu)$ (see the dashed line in Fig~\ref{fig:basicphasediagram}). This is very suggestive that the non-linear sigma model description of the problem does not apply until the disorder is strong enough to induce a relatively large zero energy DOS.
 
\begin{figure}
  \includegraphics[width=1.0\linewidth]{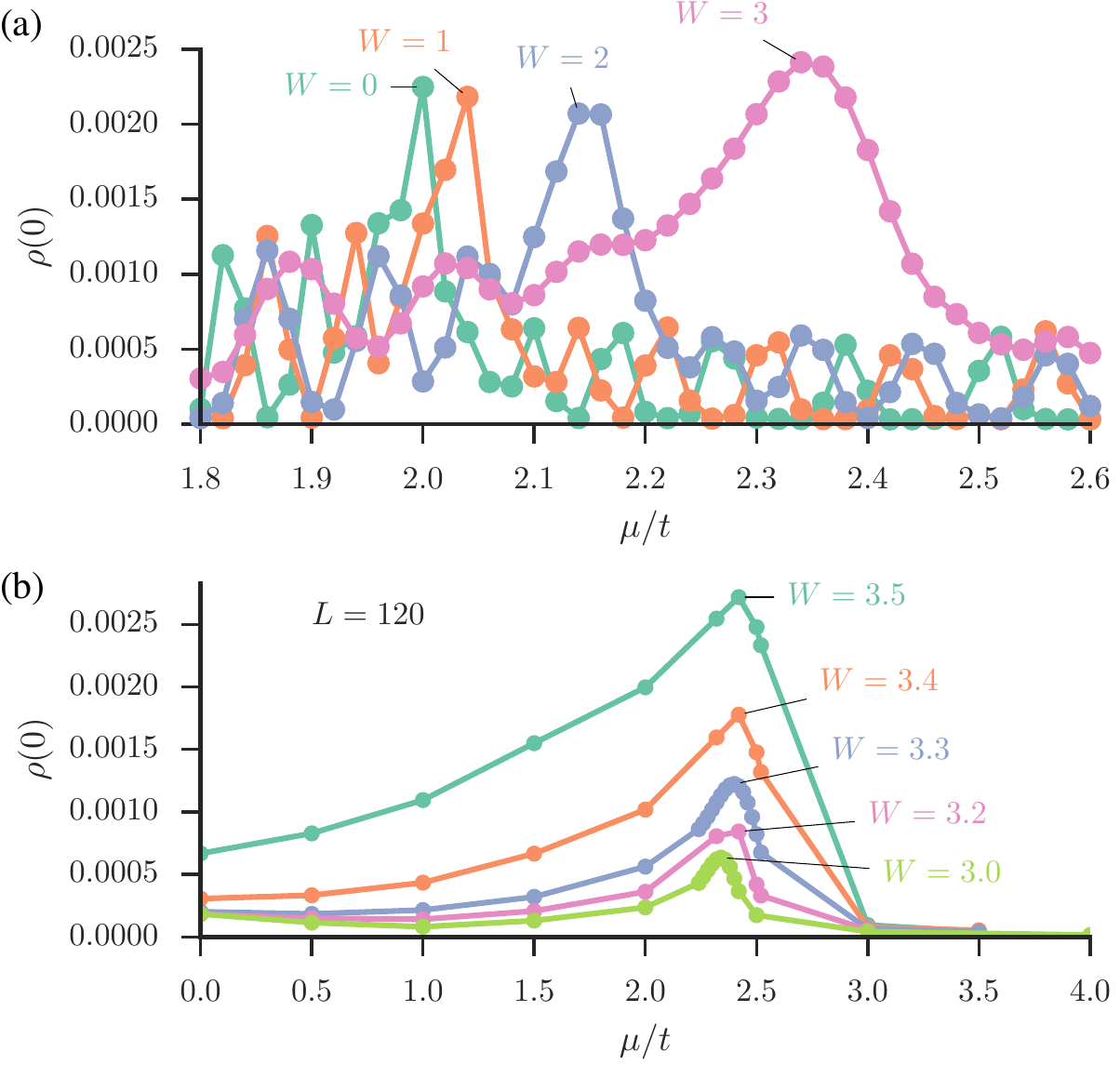}
  \caption{(a) The crossover from ThSM4 (left of peak) to ThSM2 (right of peak) by observing the finite-size induced peak for $\rho(0)$ as a function of $\mu$.
 Notice that away from the peak the data becomes noisy.
 This is the expected behavior for $\rho(0)$ in the semi-metallic regime at finite size [see Fig~\ref{fig:cleanPhaseDiagram}(c)]. 
This data is taken at $L=60$. An estimate for the full-width at half-maximum is used for the error. 
(b) The avoided multi-critical point can be visually captured here by considering $\rho(0)$ as a function of $\mu/t$. 
The peak that began at $\mu=2t$ for $W=0$ smoothly transitions to the peak seen here around $\mu\sim2.4t$. 
On the right side, we have either ThDM or ThSM4 depending on the value of $W/t$. 
On the right side, it quickly becomes ThSM2 }
\label{fig:rho0_W2_thSMthSM}
\end{figure}

In order to distinguish the two ThSM regimes at finite energy, we follow the chemical potential at which the anistropic Weyl nodes occur at finite disorder (see Section~\ref{sec:thsm4-thsm2} below), and thus identify the crossover between the two ThSM regimes $W_{\mathrm{SM}}(\mu)$. This is shown clearly in Fig.~\ref{fig:rho0_W2_thSMthSM}.

\begin{figure}
  \includegraphics[width=\columnwidth]{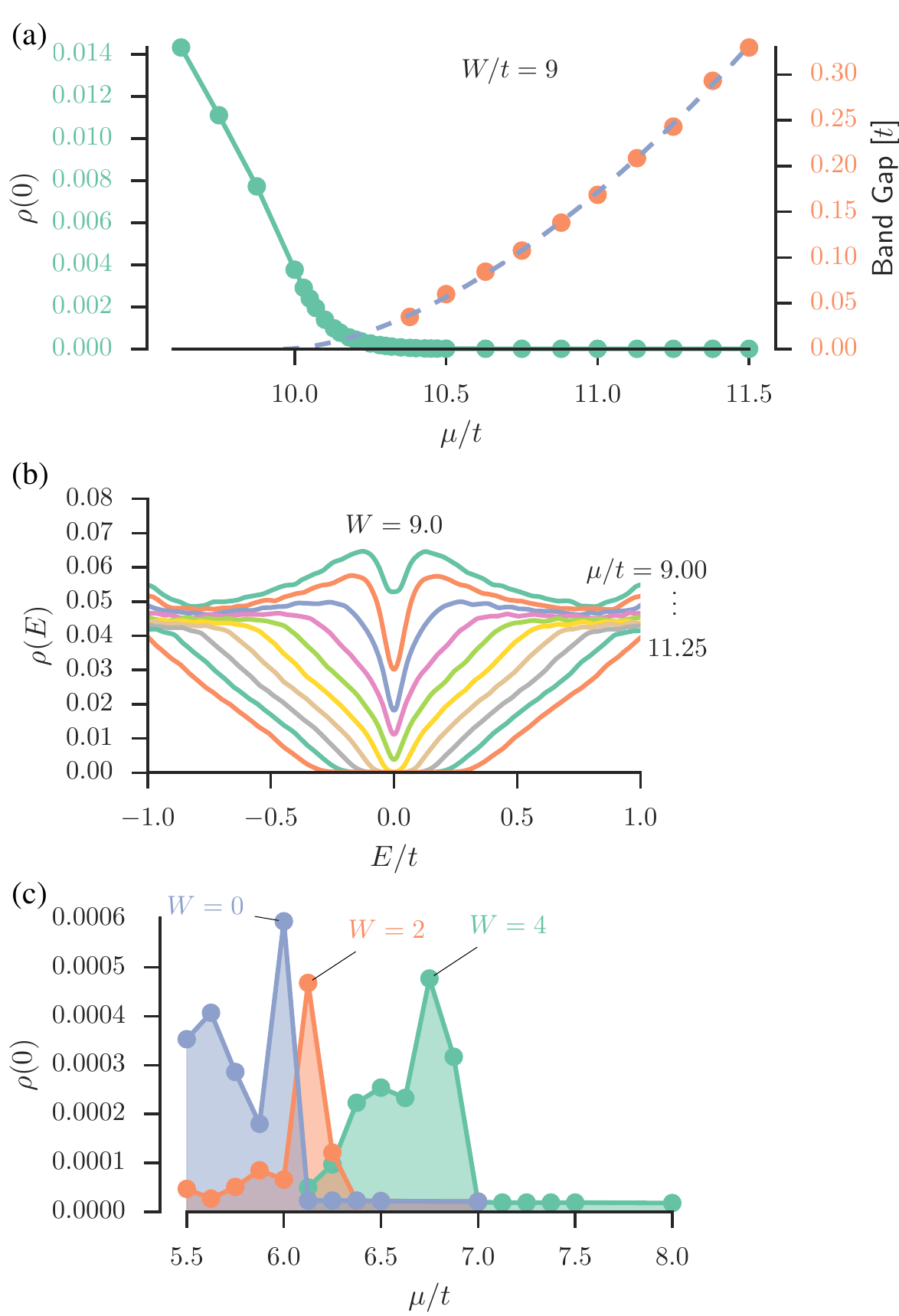}
  \caption{
  The transition from ThDM to ThBI.
 (a) $\rho(0)$ as it tends to zero from the ThDM to the ThBI (left) and the average gap that forms in the insulator (right). 
The dashed line represents a power-law fit of the gap for $L=120$.
(b) The density of states across the transition. 
Notice that roughly when the gap closes, $\rho(0)>0$ indicating the ThDM phase.
(c) The transition from the ThSM2 regime (left of peak) to the ThBI phase (right of peak) up to disorder $W/t = 4$. 
As with the ThSM4 to ThSM2 crossover in Fig.~\ref{fig:rho0_W2_thSMthSM}, the peak at the anisotropic Weyl node is expected and the subsequent fall to $\rho(0) = 0$ even at this finite size of $L=60$ is expected in the ThBI phase. 
The transition and error bars are read off the plot by the position of the peak and resolution of the data.
\label{fig:DOSW6plots}} 
\end{figure}

\begin{figure}
  \includegraphics[width=0.75\columnwidth]{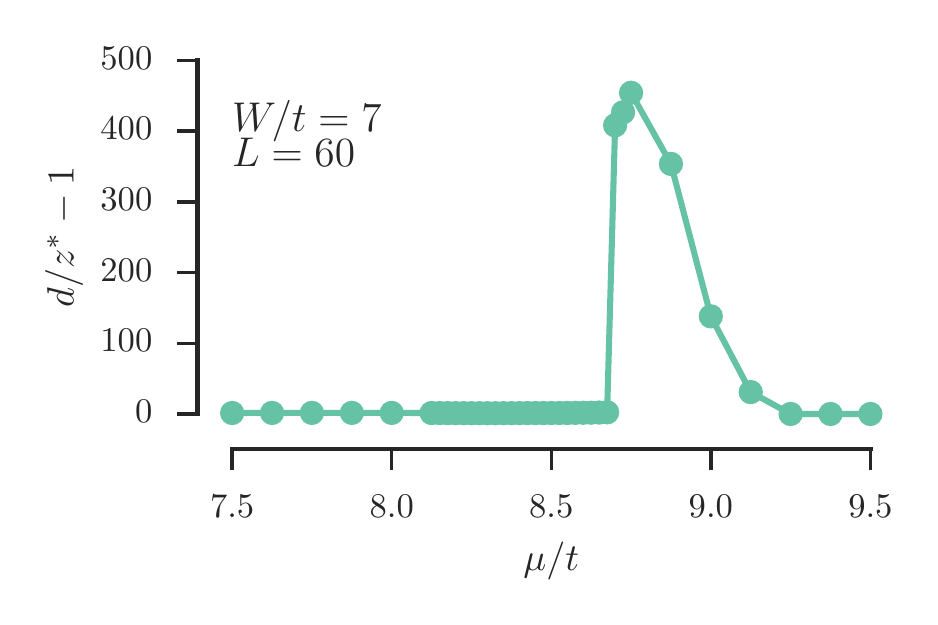}
  \caption{When fitting $\rho(E) \sim E^{d/z^*-1}$ across the phase transition from ThSM2 to ThBI, we can observe a sudden increase in the power $d/z^*-1$ due to entering the ThBI phase. This can be used to numerically give an upperbound to the transition $\mu_I(W)$.\label{fig:TransitiontoBIW7}}
\end{figure}

For fixed disorder strength below the ThDM regime (i.e.~$W<W_c(\mu)$), upon tuning $\mu$ the BdG quasiparticles become gapped out entering a ThBI phase, see Fig.~\ref{fig:DOSW6plots}.
Even considering rare regions, $\rho(0)$ is technically not a good order parameter on either side of this transition as $L\rightarrow\infty$, but its behavior at finite size allows us to nonetheless characterize the transition. In addition, the low energy power law of the DOS upon entering the ThBI phase clearly indicates the existence of an average band gap, as shown in Fig.~\ref{fig:TransitiontoBIW7}.
As discussed previously in section~\ref{sec:non-pert}, we expect Lifshitz tails should fill in a small $\rho(0)$, making $\rho(0)>0$ on both sides of the transition; however, they are Anderson localized states in the band insulator.
Separating these two phases is a QCP between a delocalized and localized phase, where in the clean limit the band structure has an anistropic Weyl point. This transition line clearly evolves under disorder and chemical potential defining the ThBI critical line $W_I(\mu)$, and we connect this transition to its clean counterpart in Section~\ref{sec:thsm-thbi}.

We discuss how each of these crossovers and transitions is determined from the data in Appendix~\ref{sec:AppendixA} for the average density of states. We discuss the thermal Anderson localization physics at much larger disorder strength separately in Section~\ref{sec:localization}.

\section{Avoided Quantum Criticality}
\label{sec:qpt}

\subsection{ThSM to ThDM crossover}
\begin{figure*}[!t]
 \includegraphics[width=2\columnwidth]{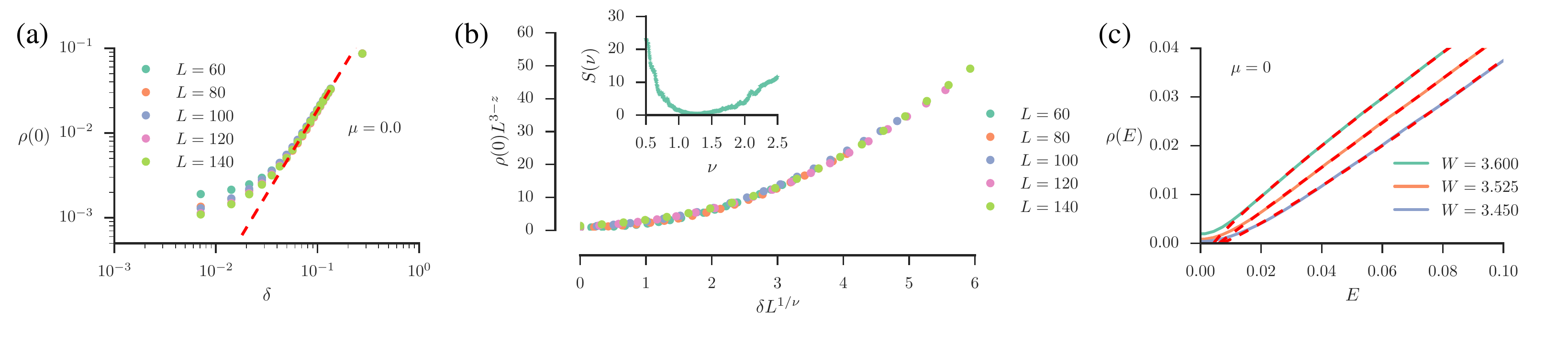}
  \caption{
    In this figure, we determine the critical exponents for the quantum critical region between the ThSM4 to ThDM regimes. 
    The exponent $\nu$ is determined by two methods: 
    (a) by $\rho(0) \propto \delta^{\nu(d-z)}$ where $\delta = (W - W_c)/W_c$ and (b) by data collapse minimizing the collapse function $S(\nu)$.
    This is illustrated here for $W_c = 3.525$.
    The exponent $z$ is found simply in (c) by fitting $\rho(E) \propto (E - b)^{\frac{3}{z} - 1}$.
    The dashed lines are numerical fits.
    The computed values for $\nu$ and $z$ are quoted in Table~\ref{table:ThSMtoDM}.\label{fig:critExpsMu0}} %fig7
\end{figure*}

For $\mu=0$ we find a crossover from ThSM4 to ThDM, which is captured in Fig.~\ref{fig:mu0SMtoDM} (a) as $\rho(0)$ becomes finite. 
As discussed in Section~\ref{sec:phase}, we find that the avoided quantum critical point is at $W_c(\mu=0)/t = 3.525 \pm 0.075$, in the inset to Fig.~\ref{fig:mu0SMtoDM} (a).
We can also see how the density of states $\rho(E)$ itself changes across the transition, as seen in Fig.~\ref{fig:mu0SMtoDM} (b).
Near $W=W_c$ we find the DOS for $E>E^*$ varies like
\begin{align}
\rho(E) \sim |E|^{d/z-1}
\end{align}
One can qualitatively see the that change from $\rho(E) \sim E^2$ to $\rho(E) \sim |E|$ occurs as a function of $W$ on a linear scale in Fig.~\ref{fig:mu0SMtoDM} (b).
with a critical exponent $z = 1.50\pm 0.09$ [as shown in Fig.~\ref{fig:critExpsMu0} (c)]. This result is in excellent agreement with the field theoretic one loop calculation for a random axial chemical potential~\cite{Goswami2011}. 
As we show in Appendix~\ref{sec:an}, various disorder couplings are generated in the RG process that are not present in the bare model. When these are all taken into account, we find that the avoided QCP in this particle-hole symmetric model is in fact dictated by the universality class of  random axial chemical potential~\cite{Goswami2011}. 
Thus, we find that the RG provides an accurate prediction for the finite energy power law scaling of the DOS. The agreement between the non-trivial power law scaling in the data and the one loop RG estimates of the critical exponents lead us to identify this as the quantum critical regime.

\begin{table}[]
\begin{tabular}{ c | c | c | c | c }
\hline \hline
$\mu/t$ & $W_c/t$           & $z$           & $\nu_{\mathrm{Fit}}$ & $\nu_{\mathrm{DC}}$ \\ \hline
0.000   & $3.525\pm0.075$   & $1.50\pm0.09$ & $1.34\pm0.22$        & $1.24\pm0.28$       \\
2.320   & $3.300\pm0.150$   & $1.50\pm0.10$ & $1.63\pm0.29$        & $1.75\pm0.16$       \\
2.420   & $3.375\pm0.150$   & $1.52\pm0.14$ & $1.42\pm0.26$        & $1.38\pm0.31$       \\
2.520   & $3.425\pm0.150$   & $1.51\pm0.10$ & $1.37\pm0.25$        & $1.29\pm0.30$       \\
5.000   & $5.275\pm0.050^*$ & $1.44\pm0.08$ & $0.98\pm0.24$        & $0.90\pm0.13$       \\ \hline\hline   
\end{tabular}
\caption{Avoided quantum critical points and critical exponents for the crossover from ThSM to ThDM.
$^*$This point was found just by fitting a power law of $\rho(0)\sim|W-W_c|^b$ on the ThDM side; it is less reliable as a result.\label{table:ThSMtoDM}}
\end{table}

We now come to numerically determining the correlation length exponent $\nu$. 
We have pointed out in Ref.~\cite{Pixley2015} that computing $\nu$ from the KPM calculation of $\rho$ suffers from large fluctuations due to the accuracy problem in $W_c$ and the size of the critical region (which is hard to determine as it depends sensitively on the strength of the avoidance). 
Despite this we can still provide a reasonable estimate of $\nu$ from the data via the power law dependence and finite size scaling similar to the method described in detail in Refs.~\cite{Pixley2015-prl,Pixley2015}.
Due to the diverging correlation length at the transition $\xi \sim |\delta|^{-\nu}$ (where we have defined the distance to the avoided critical point by $\delta = (W-W_c)/W_c$) we expect the scaling hypothesis to hold, which implies $\rho(0) \propto \delta^{\nu(d-z)}$ with $\delta>0$ for sufficiently large $L$.
Here, it is important to note that this power law dependence exists on top of the exponentially small rare region contribution, and strictly speaking $\rho(0)\neq0$ for $W=W_c$ in the thermodynamic limit.
Fitting $\rho(0) \propto \delta^{\nu(d-z)}$ as shown in Fig.~\ref{fig:critExpsMu0} (a) (using the value of $W_c$ and $z$ already determined) we find $\nu_{\mathrm{Fit}} = 1.38\pm 0.38$.
The scaling hypothesis implies the finite size scaling form $\rho(0) = L^{z-d} f(\delta L^{1/\nu})$, which we use to perform data collapse yielding $\nu_{\mathrm{DC}} = 1.24\pm0.20$, as shown in Fig.~\ref{fig:critExpsMu0} (b). 
The two values of $\nu$ are in relatively good agreement.

Similarly, we consider the crossover from ThSM2 to ThDM for various different values of $\mu$ (not shown)
with the results for the avoided critical points and their corresponding critical exponents in the quantum critical crossover regime summarized in Table~\ref{table:ThSMtoDM}.

\begin{figure}
  \includegraphics[width=\columnwidth]{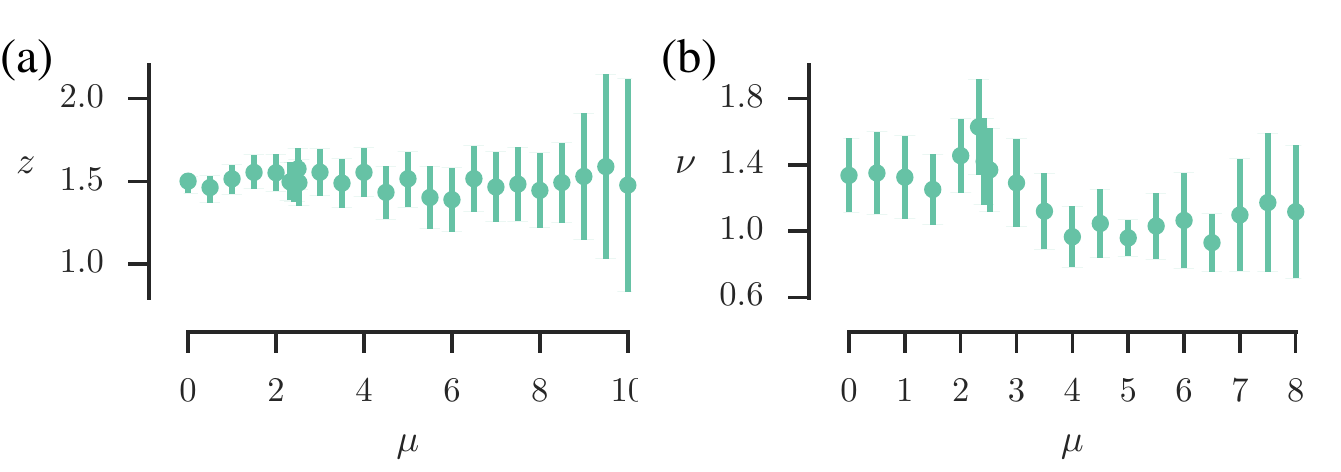}
  \caption{The critical exponents $z$ (a) and $\nu$ (b) along the critical line connecting the ThSM phases to the ThDM. It becomes more difficult to pin down a value of $z$ the closer we get to the ThBI phase. In fact, the uncertainty in $z$ is related to our inability to pin down the phase transition to a good accuracy in those regions.
  Similarly, it becomes difficult to pin down $\nu$ (when $\mu>8t$, $\nu$ cannot even be determined reliably due to proximity to the transition to ThBI). The $\nu$ data was found by finding the inflection point of $\log \rho(0)$ vs. $\log \delta$ by cubic interpolation and determining the tangent at that point. 
  Most of this data was found with $\rho(E)$ computed with $L=120$ (for $\mu=0$, $2.32$, $2.42$, $2.52$, and $5$, $L=140$ data is used). \label{fig:critExpsThSMtoThDM}} 
  \end{figure}

Lastly, we have also estimated $z$ and $\nu$ for a larger number of points focusing on $L=120$ along the avoided quantum critical line $W_c(\mu)$ as depicted in Fig.~\ref{fig:critExpsThSMtoThDM}. 
Interestingly, we find that $z \approx 1.5$ holds along the entire line.
Focusing on the avoided multi-critical point where the avoided critical line $W_c(\mu)$ [separating the ThSM regime and the ThDM regime] intersects the 
crossover line $W_{\mathrm{SM}}(\mu)$ [separating the ThSM4 and ThSM2 regimes], despite the uncertainty in $\nu$, we find a systematic increase in the value of $\nu$ near the multi-critical point.  
From a one-loop RG calculation at this multi critical point there are two relevant scaling variables and as a result $\nu_{\mathrm{MC}}=2$, which is in reasonable agreement with the numerics.

\subsection{ThSM4 to ThSM2 crossover}
\label{sec:thsm4-thsm2}

\begin{figure*}
  \centering
  \includegraphics[width=2\columnwidth]{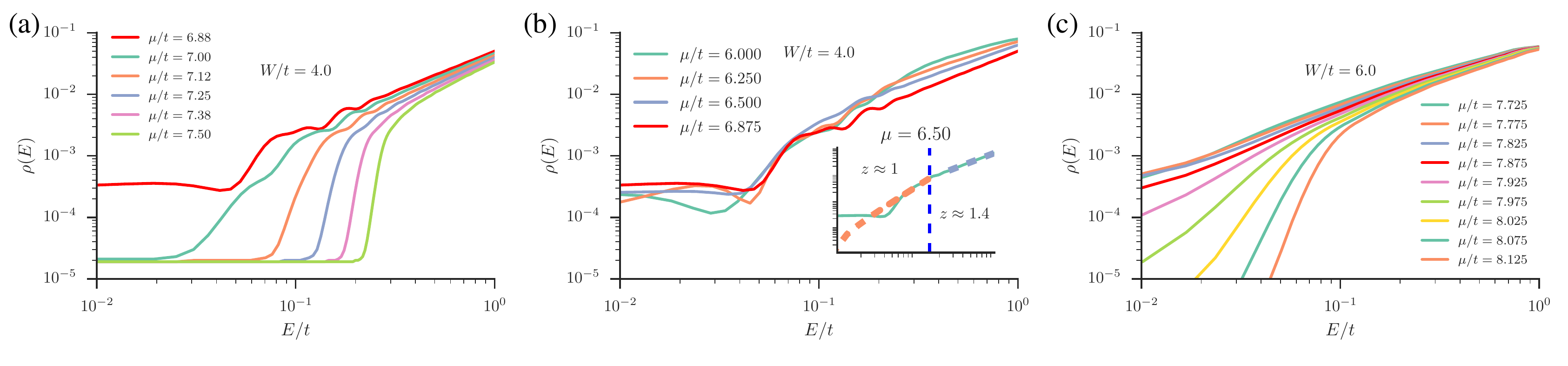} 
  \caption{The QCP associated with the ThBI to ThDM transition has a critical fan that can be observed here in the DOS.
 (a) Moving from the ThBI to the critical point (represented by red), we see the fan sharply as the gap closes.
    (b) On the ThSM side of the transition, the fan is a bit more subtle since the critical fan is a slight change in exponent $\rho(E)\sim E^2$ (or $z=1$) for the ThSM and $\rho(E)\sim E^{1.12}$ (or $z=1.4$) for the other (see the inset).
    The critical fan appears at higher energies the farther from the transition we go.
   (c) At higher disorder, the critical fan from the avoided QCP swamps all effects and we begin seeing that quantum critical fan appear.
  For all plots the red line represents the approximate critical point and we have taken $L=60$ and box potential disorder for these plots.}
  \label{fig:CriticalFanQCP}
\end{figure*}

We now turn to the crossover from one thermally semimetallic regime (ThSM4) to the other (ThSM2).
The starting point for this transition is clearly given by the peak $\mu/t=2$ in Fig.~\ref{fig:cleanPhaseDiagram}(c).
This peak can be followed to finite disorder and finally through the ThDM crossover as illustrated in Fig.~\ref{fig:rho0_W2_thSMthSM}(a) and (b).

\begin{table}[]
  \begin{tabular}{ c | c | c }
  \hline \hline
  $W/t$ & $\mu_{\mathrm{SM}}/t$  & $z_{\mathrm{SM}}^*$ \\ \hline
  0     & $2$                    & $1.2$               \\
  1.0   & $2.04\pm0.04$          & $1.22\pm0.07$       \\
  2.0   & $2.15\pm0.04$          & $1.27\pm0.09$       \\
  3.0   & $2.34\pm0.06$          & $1.33\pm0.03$       \\
  3.3   & $2.40\pm0.12$          & $1.47\pm0.07$       \\ \hline\hline   
\end{tabular}
\caption{The values of the crossover point $\mu_{\mathrm{SM}}(W)$ for various $W$ values for the crossover ThSM4 to ThSM2. Additionally, the exponent $z_{\mathrm{SM}}^*$ that describes $\rho(E)\sim E^{d/z_{\mathrm{SM}}^*-1}$ is shown. The value at $W=0$ is the clean value that is analytically known; the rest are calculated numerically.\label{table:SM4to2}}
\end{table}

In the clean limit, there are 3 anisotropic Weyl points (linear in $k_x$ and $k_y$ and parabolic in $k_z$), and we know that just as for the ThSM2 to ThBI transition, $\rho(E)\sim E^{3/2}$ or $z_{\mathrm{SM}}^* = 1.2$ by considering again
\begin{align}
  \rho(E)\sim E^{d/z_{\mathrm{SM}}^* -1}.
\end{align}
When we follow this to higher energies we get the values enumerated in Table~\ref{table:SM4to2}.
We see that the value $z_\mathrm{SM}^* \approx 1.2$ is valid until we start getting close to the transition to the ThDM at which point, we obtain results consistent with the value of $z = 1.5$. The DOS for small disorder ($W/t=1$ and $W/t=2$ in particular) 
have a large finite size effect from the Weyl peaks.
Thus, to obtain error bars (for fixed $L$) we perform a fit on the systematically-noisy data that we obtain by performing a moving average.

\section{ThDM to ThBI quantum phase transtion}
\label{sec:thsm-thbi}
\begin{figure}
 \includegraphics[width=\columnwidth]{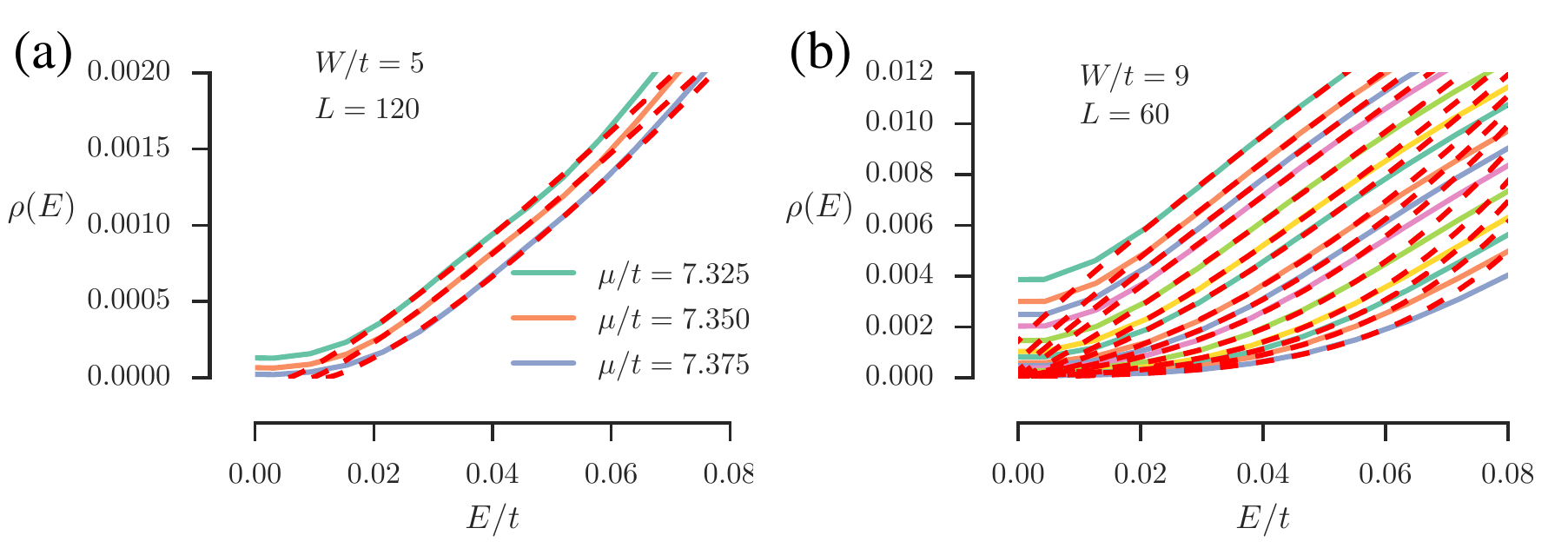}
  \caption{The fitting of the DOS at the transition from the ThSM2 regime to the ThBI phase. 
In (a) we have a smaller uncertainty for the transition, and in (b) we can fit a powerlaw within the entire region to get a smooth change from $z^*\approx 1.6$ to $z^*\approx 0.4$ within this window.\label{fig:zstarfits}}
\end{figure}
We now turn to the diffusive metal to ThBI \emph{transition}---which can be accessed at both weak and strong disorder.
In this section, we focus on the evolution of the clean anisotropic QCP (at $W=0$ and $\mu=\pm 6t$) in the presence of disorder. We consider the DOS and average band gap to see how effects of the clean QCP survive in the presence of disorder. Despite, focusing on self-averaging quantities we
are able to study the ThSM to ThBI transition (see Fig.~\ref{fig:basicphasediagram}) from both sides of $\mu_I(W)$ using $\rho(E)$ and the average gap $\Delta_g$. 
Note that we will use the notation for the transition line $\mu_I(W)$ and its functional inverse $W_I(\mu)$ interchangeably. 

\begin{figure}[!ht]
  \begin{minipage}{0.45\textwidth}
  \includegraphics[width=\columnwidth]{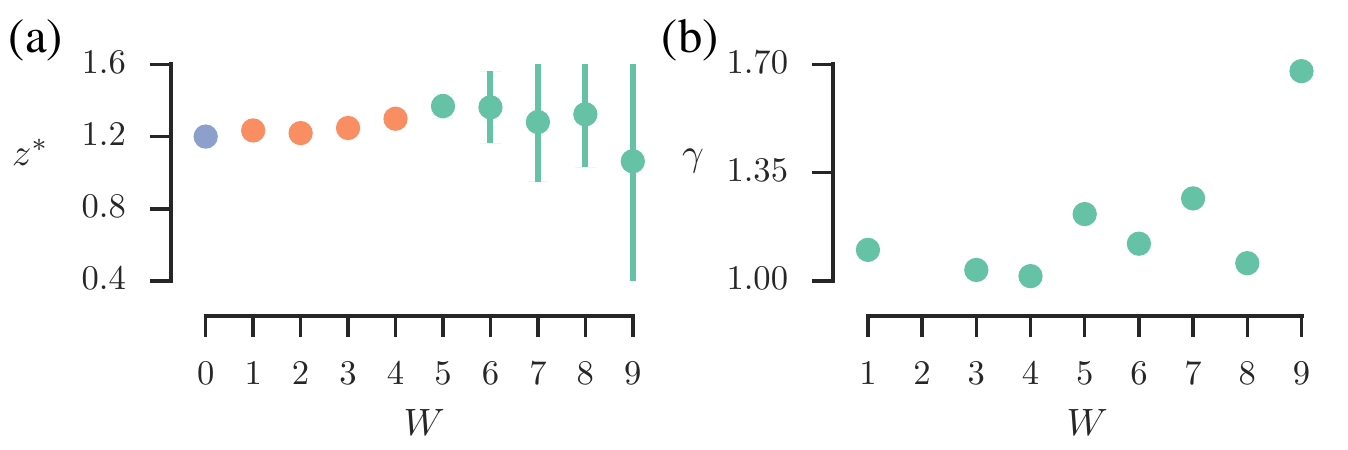}
  \end{minipage}
  \caption{The critical exponents on the critical line $\mu_I(W)$. 
(a) The critical exponent $z^*$ extracted from fitting $\rho(E) \sim E^{d/z^* -1}$. 
In the ThBI phase $z^* \rightarrow 0^+$.
For $W=0$ (the blue data point), this value is known exactly from analytics. 
For $1\leq W\leq 4$ (the orange data points), $\rho(E)$ is noisy, so we find this exponent for larger bounds that could see higher band effects. 
For $W\geq 5$, we can fit $\rho(E)$ with $E/t\ll 1$ and these are the exponents we find.
(b) The critical exponent  $\gamma=\nu z$, that describes how the band insulator the gap increases in the ThBI phase: $\Delta_g \sim (\mu -\mu_c)^\gamma$. 
Systematics that could lead to appreciable error in this quantity cannot be reliably estimated. 
  \label{fig:critExpThBI}} 
\end{figure}

We are able to connect this QCP to the clean limit by studying the energy dependence of the density of states. Along the critical line $W_I(\mu)$ we compute the dynamic exponent $z^*$ from 
\begin{align}
\rho(E) \sim E^{d/z^* -1}.
\end{align}
We show the corresponding numerical results for $z^*$ in Fig.~\ref{fig:critExpThBI} (a), obtained from the fitting shown in Fig.~\ref{fig:zstarfits}.  
We find good agreement with the dynamic exponent in the clean limit ($=1.2$) along $W_I(\mu)$ until $W_I(\mu)\gtrapprox 7.0t$, which is in good agreement with the expectation that disorder acts like an irrelevant \emph{perturbation} to the anisotropic QCP.
As we have shown in Section~\ref{subsec:thdmtobi}, non-perturbative effects of disorder round out this non-analyticity in the DOS at the lowest energy (or longest length scale), and the quantum critical scaling only holds at finite energy above a non-universal cross over scale. Similar to the avoided QCP, due to the good agreement between the numerics and the analytic estimate of critical exponents from the one loop RG, we associate this with the quantum critical regime. 
When $W_I(\mu)\gtrapprox 7.0t$, the quantum critical fan from the avoided quantum critical point begins to contaminate the calculation for $z^*$, leaving us with a wide range of exponents and large error bars. Our KPM numerics are no longer reliable for determining critical exponents in this regime.

Upon entering the ThBI phase an average band gap opens near zero energy, which is captured in $\rho(E)$ depicted in Fig.~\ref{fig:DOSW6plots}(b). This allows us to determine the average band gap in the ThBI phase. 
In both cases, we also use the vanishing of gap as an order parameter for the transition.     
Approaching $\mu_I(W)$ from the ThBI side we find that the average gap vanishes in a power law fashion 
\begin{align}
\Delta_g \sim |\mu-\mu_I(W)|^{\gamma},
\end{align}
where $\gamma = \nu z$ since $\Delta_g\sim \xi^{-z}$ with a correlation length $\xi \sim |\mu - \mu_I(W)|^{-\nu} $.
The values of $\gamma$ are given in Fig.~\ref{fig:critExpThBI} (b) which vary between 1 and 1.3 upon entering the ThBI and jump up to 1.7 for $W_I(\mu)=9$, closer to where the disorder is so large that there is no remnant of the SM regime any longer.
Note that, since the non-analytic behavior in the DOS is rounded out, this implies that the scaling of the average band gap is also rounded on the largest length scales. 
Our data is consistent with $\nu \approx 1$ for most of the critical line if $z^*\approx 1.2$, 
until it becomes closer in proximity to the avoided QCP and $z\approx 1.5$, while $\nu$ remains roughly the same. 

The crossing of quantum critical fans can be seen in Fig.~\ref{fig:CriticalFanQCP} where the different regimes can be seen explicitly in the DOS. 
This is a verification of our schematic in Fig.~\ref{fig:phasediagram}(b) near the QCP.
For larger disorder, the ThDM regime gets so close to the ThBI phase that we can no longer properly discern a ThSM regime. 
Our results in Fig.~\ref{fig:SaturateSMtoBI}(c) are inconclusive in regards to whether the avoided QCP and the true transition merge; instead we merely see a broad and ill-defined peak in $\rho''(0)$ as a function of $\mu$. 

The critical exponents ($\nu$ and $z^*$) that we have computed in this section describe the power law scaling in the DOS and the average band gap. Our numerical estimates are in good agreement with the one loop RG calculations in Appendix~\ref{sec:an}. 
Approaching this transition from the ThBI side, the localization transition will be described by a diverging localization length $\xi_l\sim|W-W_I|^{-\nu_l}$ and will give rise to single parameter scaling in quantities that are not self averaging (such as the inverse participation ratio or the typical DOS), with robust non-analytic behavior at $E=0$. It will be interesting to study this localization transition in the future using observables that explicitly track the localization length and see if there is any relation between these two sets of exponents (e.g. $\nu$ and $\nu_l$).

\begin{figure}
  \includegraphics[width=0.4\textwidth]{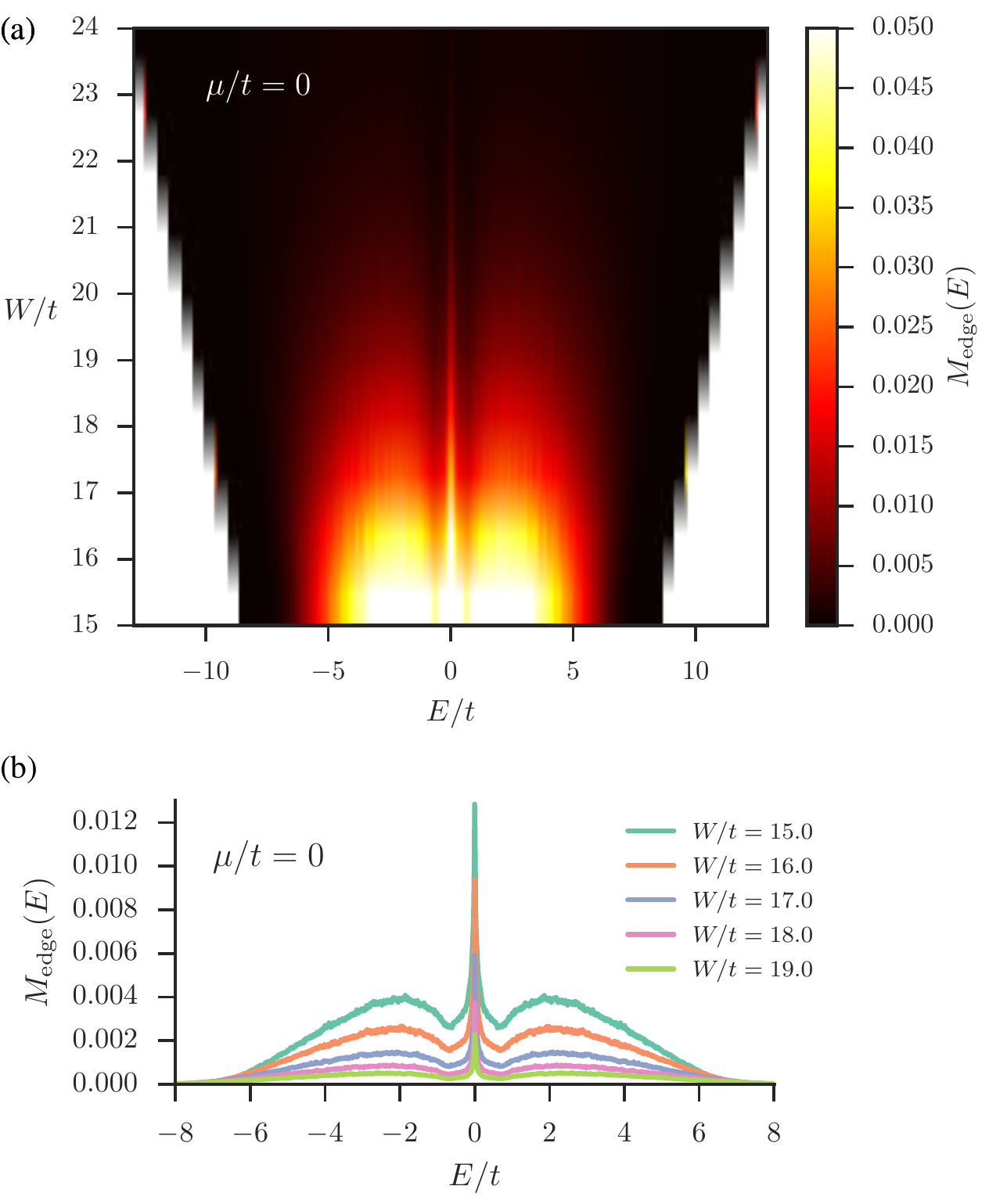}
  \caption{Plots of the mobility edge defined by $M_{\mathrm{edge}}(E) = \rho_t(E)/\rho(E)$ tuning disorder $W$ while keeping $\mu=0$ constant. 
  We see a large anti-localization peak in the DOS develop in addition to the standard behavior that states far from $E=0$ localize first. 
  The function $\rho(E)$ is calculated at $L=60$ and $\rho_t(E)$ is calculated at $L=30$ and $N_C = 2^{13}$. 
  In (a) we have the whole range given by a  color
  plot, and in (b) we see some cuts illustrating how drastic the peak is. 
  ~\label{fig:MedgeFigures0}} 
  \end{figure}

\section{Thermal Anderson Insulator and localization at large disorder}
\label{sec:localization}

We now turn to the thermal Anderson insulator properties at large disorder.
It is important to note that we are now considering a much larger disorder strength than we have considered so far. 
In order to study localization phenomenon, we consider both the typical $[\rho_t(E)]$ and the average $[\rho(E)]$ density of states just as in Sec.~\ref{sec:ThermalDiffMetal}. 
Since the typical DOS goes to zero in the thermal AI phase we can define the mobility edge, as $M_{\mathrm{edge}}(E) = \rho_t(E)/\rho(E)$.
As shown in Fig.~\ref{fig:MedgeFigures0}, 
we find a thermal mobility edge at finite energy, separating states that are a ThDM and thermal Anderson insulator.
The peak from Sec.~\ref{sec:ThermalDiffMetal} shows up here before the transition.
In addition, as shown in Fig.~\ref{fig:MedgeFigures0} (a), 
the thermal mobility edge squeezes in towards zero energy where states far away from zero energy localize first.

\begin{figure}
  \includegraphics[width=\columnwidth]{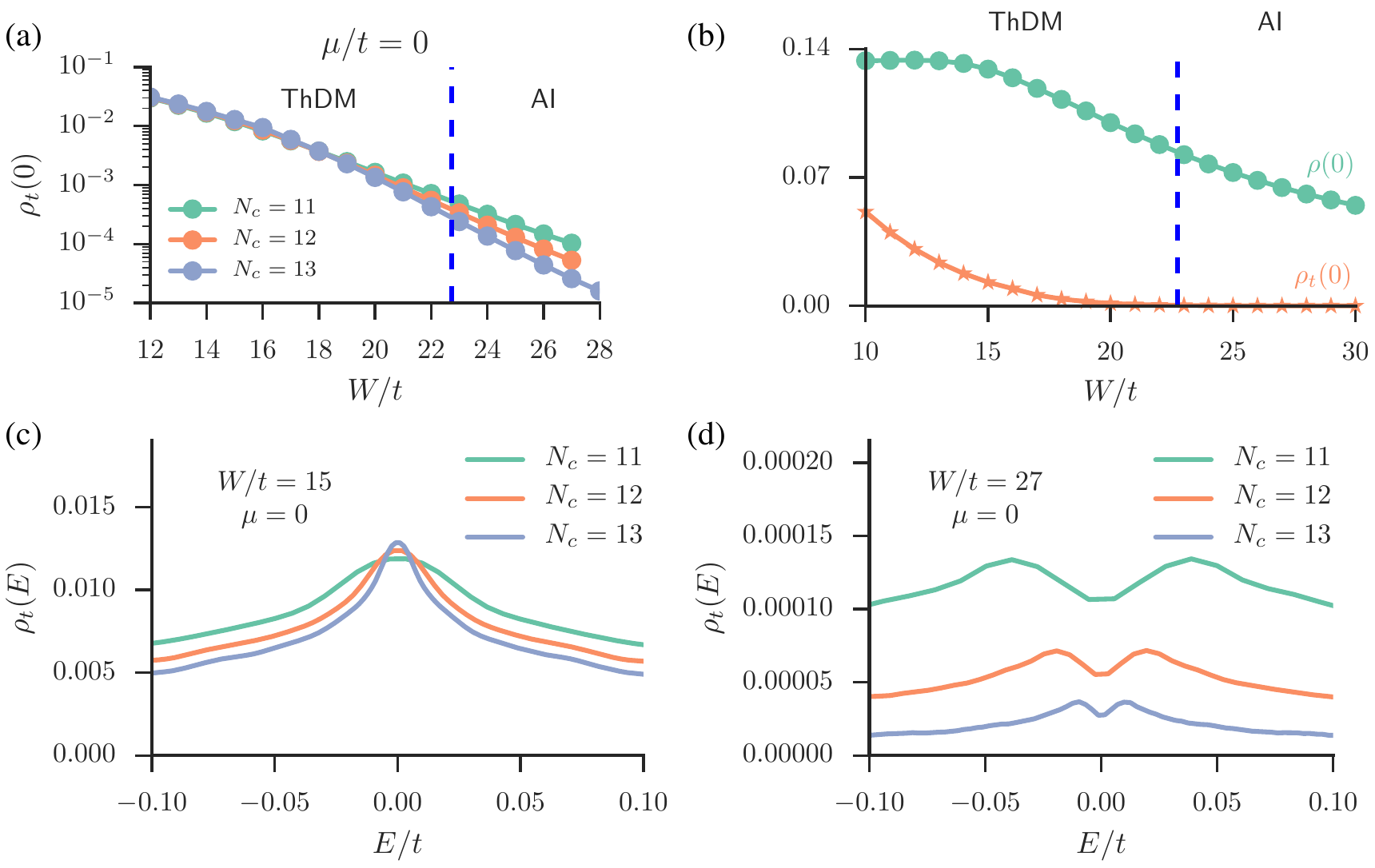}
  \caption{Typical DOS displaying the Anderson localization transition at large disorder for $\mu=0$ and $L=30$. (a) The $N_C$ dependence of the typical DOS. In the localized phase, $\rho_t(0)$ decreases as $N_C$ is increased whereas in the ThDM phase $\rho_t(0)$ is relatively insensitive to $N_C$.
    While this is plotted for $\mu=0$, the plot is nearly identical for any $\mu$ [see Fig.~\ref{fig:powerLawFitLocalization} (a)]. (b) The typical DOS goes to zero while the average DOS remains finite. The average DOS is computed with $L=60$. (c) The peak in the typical-DOS before the transition to the AI. One sees little or even positive $N_C$ dependence at $E=0$.
    (d) The peak splits and now $\rho_t(E)$ has a strong $N_C$ dependence.
    With the theory for the ThDM breaking down and other measures, we conclude this is the localized phase.
  \label{fig:AItransition}}
\end{figure}

However, in order to see localization, we need to check the $N_C$-dependence of the typical DOS $\rho_t(0)$.
The finite KPM expansion order ($N_C$) controls the broadening of the Dirac-delta function in the local DOS. This introduces an artificial length scale that can make $\rho_t(0)$ ``look'' more delocalized due to the convolution of the single particle wave functions and the broadened Dirac-delta functions.
For increasing $N_C$ this length scale in $\rho_t(0)$ should vanish as $N_C\rightarrow\infty$, and thus in the localized phase $\rho_t(0)$ should go to zero with $N_C\rightarrow \infty$.
As shown in Fig.~\ref{fig:AItransition} (a) we find the transition point by fitting the data at each $N_C$ to a power law $\rho_t(0)\sim(W_l - W)^\beta$ and then fitting $W_l$ to a polynomial in $N_C$, $W_l = a/N_C^2 + b/N_C + c$.
We then extrapolate $W_l$ to $N_c\rightarrow \infty$ to obtain the true transition point.
This is represented in Fig.~\ref{fig:AItransition} (b).

However, one might wonder what happens to the peak in the average and typical DOS in the AI phase.
The peak in the average gets broadened [see Fig.~\ref{fig:peaks} (b)], but the peak in the typical DOS behaves very differently.
Before the transition, one can see that the peak is more or less intact in Fig.~\ref{fig:AItransition} (c), but after the transition to the AI, the peak splits and gets a strong $N_C$-dependence as seen in Fig.~\ref{fig:AItransition} (d).

\begin{figure}
  \centering
  \includegraphics[width=\columnwidth]{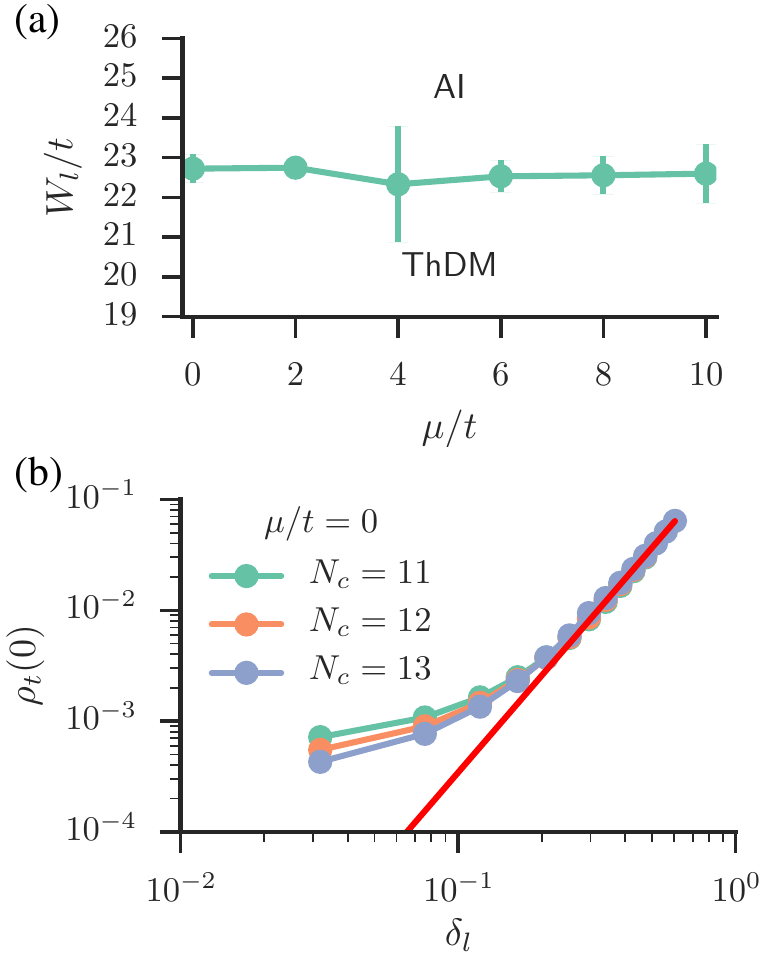}
  \caption{(a) The Anderson insulator transition line at large disorder extracted from the $N_C$ dependence of the typical DOS.  (b) The fit of the power law for the typical DOS across the Anderson localization transition in the regime that is $N_C$ independent. We find $\beta = 2.9\pm 0.1$ and the fit is the red line. All data here is taken at $L=30$.
  \label{fig:powerLawFitLocalization}}
\end{figure}

Since there is a localization phase transition, we can also characterize the critical exponent by $\rho_t(0)\sim \delta_l^\beta$ where $\delta_l = |W_l-W|/W_l$. 
However, despite the uniformity of the data for all $\mu$, the slowly encroaching phase transition for large $\mu$ modifies this behavior dramatically and leads to what appears to be a lowering of $\beta$ for larger $\mu$.
We cannot confidently say that this is a real affect on the critical exponent $\beta$, so we merely state its value at $\mu=0$ which is $\beta = 2.9\pm0.1$ and the fit can be seen in Fig.~\ref{fig:powerLawFitLocalization} (b).
However, $\rho_t(0)$ and $\rho(0)$ seem rather $\mu$-independent at around $W\sim 22t$ where the Anderson transition occurs.

\section{Discussion and Conclusion}
\label{sec:conclusions}

In this work we have shown that three-dimensional disordered spinless $p_x+ip_y$ superconductors have a rich quantum phase diagram with various types of crossovers, non-perturbative effects of disorder, quantum phase transitions between different superconducting phases, and Anderson localization. 
This opens the door for the study of disordered three-dimensional nodal superconductors in general.  
Our work can also be thought of as the general theory for the consideration of quantum phases in Majorana-Weyl fermions in the presence of quenched disorder.

We have established the existence of non-perturbative quasi-localized rare states in the presence of particle-hole symmetry.
(It is important to emphasize here that the earlier extensive work~\cite{Pixley2016,Pixley2016a} in this context on nonperturbative rare region effects by two of the coauthors were restricted to systems without (in italics) particle-hole symmetery.) 
This finding is highly non-trivial as prior to our work it was natural to assume that particle hole symmetry could somehow ``protect'' the zero energy eigenstates, but we have found that this is not true. 
As a result of disorder-induced rare regions all three-dimensional nodal superconductors will always have a non-vanishing DOS at the Fermi energy. 
However, this effect is exponentially small in disorder, and may be therefore difficult to detect experimentally (but perhaps no more difficult than in the corresponding non-superconducting systems~\cite{Pixley2016,Pixley2016a}). 
Nonetheless, we do expect that the avoided quantum critical fan can be probed in thermodynamic quantities such as the specific heat or even the thermal conductivity. 
We therefore expect that our results will be particularly relevant to situations involving doping of various heavy fermion superconductors. 
Our main theoretical accomplishments are: (1) showing that nonperturbative rare region effects convert various disorder-driven `semimetal' to `diffusive metal' transitions to crossovers with avoided criticality (with the system being always a diffusive metal even at very weak disorder in contrast to the perturbative RG theory); (2) the underlying avoided critical physics can be well-described by a one-loop RG calculation with reasonable agreement  between the RG theory and exact numerical calculations; (3) non-perturbative effects of disorder can round out non-analytic behavior in the clean DOS; (4) at strong disorder, the three-dimensional class D diffusive metal phase undergoes an Anderson localization transition to an Anderson insulator.
We emphasize that our work now establishes these conclusions to be definitive for systems with particle-hole symmetry with earlier work~\cite{Pixley2016,Pixley2016a} establishing it without this symmetry.  Particle-hole symmetry allows additional phases and phase transitions in the system which were not considered before.

A new feature of the model we have considered here, that is distinct from our previous work on Dirac and Weyl semimetals (Refs.~\cite{Pixley2015-prl,Pixley2015,Pixley2016,Pixley2016a}), is the presence of a (thermal) semimetal to band insulator QCP in the band structure. At this transition, the single particle dispersion still has nodal points but the scaling is anisotropic. This leads to a non-analytic DOS that vanishes like $\rho(E)\sim|E|^{3/2}$, and represents a distinct phase from that of the semimetal. Our RG results predict that disorder acts as an irrelevant perturbation at this clean fixed point, and the non-analytic behavior should hold along the renormalized phase boundary (in disorder and chemical potential). As we have shown, however, non-perturbative effects of disorder round this out and the DOS becomes analytic across the entire phase diagram. The explicit theory of these anisotropic rare states is unknown at present and is an interesting problem for future work.
 Nonetheless, our results point to the generic scenario that non-perturbative effects of disorder dominate the generic behavior of three-dimensional Fermi points (at the longest length scales) independent of symmetry classifications. Disorder acts like an irrelevant perturbation making this class of problems distinct from those with a DOS that does not vanish faster than $|E|$ (e.g. graphene or two-dimensional $d$-wave superconductors), which is an important new distinction arising from our work.

It will be important in the future to treat the superconducting phase in a self consistent manner. 
Here we described phases, crossover regimes, avoided criticality, and other properties of the a nodal class-D Hamiltonian, but a real system will have  a fluctuating order parameter affected by disorder.
The existence of these rare states are likely not very sensitive to a spatially fluctuating superconducting gap. 
As a result, our work opens the prospect of finding rare region mediated superconductivity, where the quasi-localized large probability amplitude is likely to produce tunneling of the Bogoliubov-de Gennes quasiparticles between the rare regions leading to large puddles of superconductivity that will eventually become phase coherent~\cite{Nandkishore2014} at sufficiently low temperatures. 
A detailed study of this physics is well outside the scope of the current work focusing on the quantum criticality (or lack thereof), but should be an important future extension of our work.  Such a rare region mediated superconductivity will essentially be a novel and exotic phase of matter.

The phase diagrams in Fig.~\ref{fig:phasediagram} and Fig.~\ref{fig:basicphasediagram} summarize all of the physics presented in this manuscript.
Our simple model captures much of the essential physics of gapless Weyl nodes in the presence of particle-hole symmetry.
We explored much of this physics in the current work: the role of rare-regions, the nontrivial density of states in the diffusive metal, and the localization transition at higher values of disorder.
Despite rare-regions, we were still able to probe field theoretic quantites like critical exponents near the avoided critical point, and characterize the physics near true transitions such as the thermal insulator to thermal metal transition.
Our analytical and numerical works agree well with each other where ever they both apply.  The current work along with the earlier works presented in Refs.~\cite{Pixley2015-prl,Pixley2015,Pixley2016,Pixley2016a} essentially complete the basic theoretical study of disorder-driven quantum criticality and nonperturbative rare region physics in three dimensional Weyl systems, bringing to an end a quest that started thirty years ago with Refs.~\cite{Fradkin1986,Fradkin1986a}.

\begin{acknowledgments}
We thank Olexei Motrunich, Gil Refael, Matthew Foster, Sarang Gopalakrishnan, and Rahul Nandkishore for useful discussions. We also thank David A. Huse for various discussions and collaborations on earlier work. This work is partially supported by JQI-NSF-PFC and LPS-MPO-CMTC (JHP, PG, and SDS), and the Airforce Office for Scientific Research (JHW).
 The authors acknowledge the University of Maryland supercomputing resources (http://www.it.umd.edu/hpcc) made available for conducting the research reported in this paper.
 Part of this work was performed at the Aspen Center for Physics, which is supported by National Science Foundation grant PHY-1066293. 
 \end{acknowledgments}

\appendix

\section{Perturbative renormalization group analysis} 
\label{sec:an}

In this appendix we determine the perturbative effects of disorder on the model defined in Eq.~(\ref{eq:pipHamiltonian}).
 Our goal is to establish the universality class, within one loop RG, that governs the perturbative QCPs. Despite this method missing the non-perturbative effects of disorder, the RG does provide an analytic understanding of various features we have observed in the numerics.

\subsection{Anisotropic QCP between ThSM2 and ThBI} 
We first consider the disorder effects on the single flavor of critical excitations governing the QCP between ThSM2 and ThBI. In the presence of a random chemical potential $V(\mathbf{r})$ (for normal quasiparticles) following Gaussian white noise distribution, the replicated Euclidean effective action around the critical point becomes 
\begin{eqnarray}
&&\bar{S}_c=\int d^3x d\tau \psi^\dagger_a[\partial_\tau - i v_p \nabla_\perp \cdot \boldsymbol \tau + (c_1 \partial^2_3 + \delta \mu)\tau_3]\psi_a \nonumber \\
&-&\frac{\Delta_0}{2} \int d^3x d\tau d\tau^\prime [\psi^\dagger_a \tau_3 \psi_a](\mathbf{x},\tau)[\psi^\dagger_b \tau_3 \psi_b](\mathbf{x},\tau^\prime),  
\end{eqnarray}
where $a$ and $b$ are replica indices, and $\delta \mu=6t-\mu$ is the deviation from the QCP, $v_p=\Delta_p$ is the quasiparticle velocity in the $xy$ plane, $c_1=t$ is the inverse effective mass along the $z$ direction, and $\Delta_0$ is the disorder strength.  The quadratic part of $\bar{S}_c$ remains invariant under the scale transformations $\mathbf{x}_\perp \to \mathbf{x}_\perp e^l$, $\tau \to \tau e^{l}$ and $x_3 \to x_3 e^{l/2}$, and $\psi \to \psi e^{-5l/4}$, while the disorder coupling changes according to $\Delta_0(l)=\Delta_0(0) e^{-l/2}$. Therefore, weak disorder is an irrelevant perturbation, which can only modify the location of the QCP or the phase boundary between ThSM2 and TBI. At one loop level the RG flow equation for $\Delta_0$ is given by
\begin{equation}
\frac{d\Delta_0}{dl}=-\frac{\Delta_0}{2}-\mathcal{B}_1 \Delta^2_0,
\end{equation}    
where $\mathcal{B}_1$ is a nonuniversal constant that depends on the precise method of mode elimination. Therefore, the leading order quantum corrections make disorder a more irrelevant perturbation at the QCP. This is reminiscent of mass disorder effects on a two dimensional, two component Majorana fermion that separates a thermal quantum Hall and a thermal band insulator. In fact, by setting $c_1=0$, one accesses this particular case of a two dimensional quantum phase transition. To summarize, the QCP separating ThSM2 and ThBI remains stable against disorder, even after accounting for quantum corrections at one loop level, in contrast to the predictions of SCBA. The 
RG flow equation for $\delta \mu$ is given by
\begin{equation}
\frac{d\delta \mu}{dl}=(1-\mathcal{B}_2 \Delta_0)\delta \mu + \mathcal{B}_3 c_1 \Delta_0
\end{equation}
where $\mathcal{B}_{2}$ and $\mathcal{B}_{3}$ are two additional regulator dependent, positive constants. After solving the two flow equations simultaneously, we find 
\begin{equation}
[\delta \mu(l) + \frac{2}{3} \mathcal{B}_3 c_1 \Delta_0(l)] \approx [\delta \mu(0) + \frac{2}{3} \mathcal{B}_3 c_1 \Delta_0(0)] e^l,
\end{equation}
for weak disorder. Therefore, $\delta \tilde{\mu}=[\delta \mu(0) + \frac{2}{3} \mathcal{B}_3 c_1 \Delta_0(0)]=0$ defines the renormalized phase boundary between ThSM2 and ThBI. After noting that $\delta \mu(0)=6t-\mu$, we find that disorder shifts the phase boundary to a larger value of chemical potential $\mu=t(6 +  \frac{2}{3}  \mathcal{B}_3 \Delta_0 )$, thus expanding the ThSM2 region. If $\delta \tilde{\mu} <0$, we need to integrate the flow equations up to $l_\ast=\log \left(1/|\delta \tilde{\mu}| \right)$, and work with the low energy theory of disordered MW fermions. Hence, in the vicinity of the ThSM2 to ThBI transition, $\Delta_0(l_\ast)<\Delta_0(0)$ acts as the bare disorder coupling for the MW fermions. Consequently, we expect $W_c(\mu)$ to increase when $\mu$ approaches the ThSM2 to ThBI phase boundary.

\begin{figure*}%[!b]
\centering
\includegraphics[width=2\columnwidth]{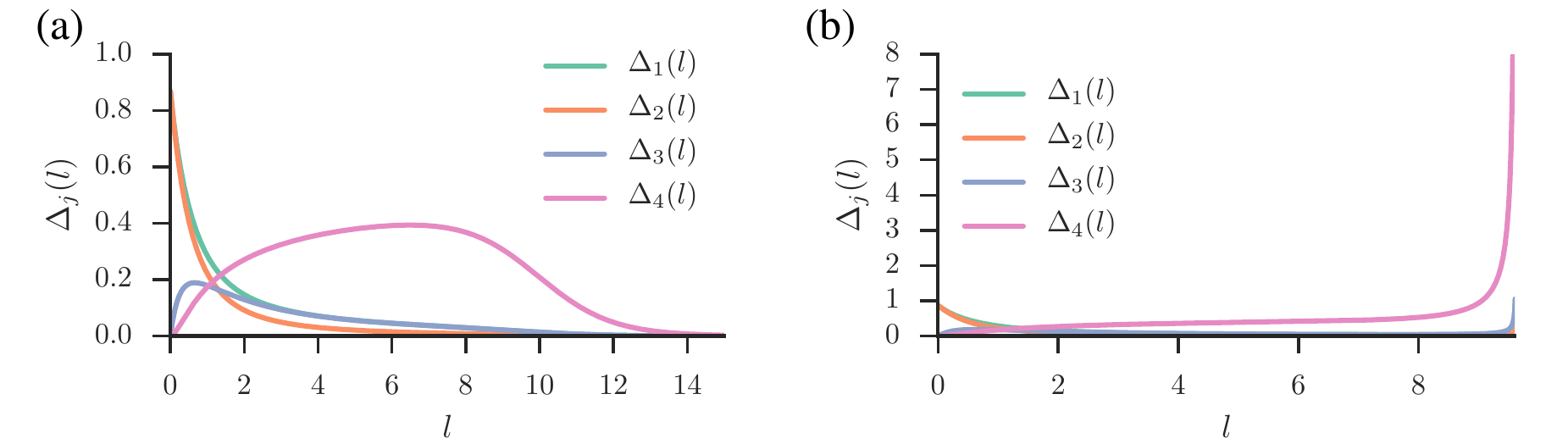}
\caption{(Color Online) Renormalized disorder couplings vs. RG flow time $l=\log(\Lambda_0/\Lambda)$. The disorder couplings $\Delta_1$, $\Delta_2$, $\Delta_3$ and $\Delta_4$ are respectively showed as the green, orange, blue and purple lines. (a) for the initial values $\Delta_1(0)=\Delta_2(0)=0.8701$, $\Delta_3(0)=\Delta_4(0)=0$ the renormalized disorder couplings flow to zero at long wavelength limit, signifying a perturbatively stable thermal semimetal phase. (b) for the initial values $\Delta_1(0)=\Delta_2(0)=0.8702$, $\Delta_3(0)=\Delta_4(0)=0$, the renormalied axial chemical potential coupling $\Delta_4(l)$ diverges at long wavelength limit, indicating a diffusive metal phase. Therefore the universality class of quantum phase transition between semimetal and diffusive metal phases is described by the axial chemical potential disorder controlled quantum critical point, even though we are explicitly tuning the strength of random chemical potential for normal quasiparticles.}    
\label{RGFlow}
\end{figure*}

\subsection{Majorana-Weyl fermions: ThSM to ThDM perturbative QCP} Inside the ThSM2 phase, the Nambu spinor $\psi(x)$ can be written in terms of the two component right and left handed MW fermion fields as $\psi(\mathbf{x}) = R(\mathbf{x}) e^{i K_1 x_3} +L(\mathbf{x}) e^{-i K_1 x_3}$. After combining the right and left handed fields $R$ and $L$ into a four component spinor $\Psi^T=(R^T, L^T\tau_3)$, the effective action for the MW fermions can be written in the following form
\begin{eqnarray}
S_{MW}&=&\int d^3x d\tau \Psi^\dagger [\partial_\tau - i v \partial_j \Gamma_j]\Psi,
\end{eqnarray}
where $\Gamma_j=\sigma_3 \otimes \tau_j$ are three anticommuting gamma matrices. We have rescaled the spatial coordinates according to $\mathbf{x}_\perp \to (v_F/v_p)^{1/3} \mathbf{x}_\perp$, $x_3 \to (v_p/v_F)^{2/3} x_3$, where $v_F=2t\sin(K_1)$ is the velocity of Weyl fermions along the $z$ or nodal direction. Consequently, we have obtained an isotropic quasiparticle velocity $v=v^{1/3}_{F} v^{2/3}_{p}$. Since the other two anticommuting gamma matrices $\Gamma_4=\eta_1 \otimes \tau_0$ and $\Gamma_5=\eta_2 \otimes \tau_0$ are absent from the effective action, the system has a continuous chiral symmetry under the operation $\Psi \to e^{i \theta \Gamma_{45}}$ where $\Gamma_{45}=-\Gamma_4 \Gamma_5=\eta_3 \otimes \tau_0$. Physically the chiral symmetry originates from the underlying translational symmetry.  

The random chemical potential for normal quasiparticles gives rise to (i) intranode scattering term of the form $V_{3,45}(\mathbf{x}) \Psi^\dagger \Gamma_3 \Gamma_{45} \Psi$, and (ii) two internode scattering terms $V_{b,1}(\mathbf{x})\Psi^\dagger \Gamma_4 \Psi$ and $V_{b,2}(\mathbf{x})\Psi^\dagger \Gamma_5 \Psi$. The intranode scattering term acts as the third component of the random axial vector potential, causing a random variation of the nodal separation along the $z$ direction. By contrast, the internode scattering terms act as random Dirac masses. We will consider these intranode and internode scattering terms as independent random variables, following Gaussian white noise distributions, and respectively assign the coupling constants $\Delta_1$, and $\Delta_{b,1}=\Delta_{b,2}=\Delta_2$. For a short range random chemical potential, the bare intranode and internode coupling constants are almost equal.

When we coarse grain the replicated, disorder averaged effective action obtained from the above random potentials, additional intranode scattering terms are generated. They are described by (i) $\sum_{j=1}^{2} V_{j,45}(\mathbf{x})\Psi^\dagger \Gamma_j \Gamma_{45}\Psi$ and (ii) $V_{45}(\mathbf{x}) \Psi^\dagger \Gamma_{45} \Psi$. At the microscopic level, the first type of scattering arises from random triplet pairing along the $z$ direction (i.e., with a form factor $\sin k_3$). For the low energy problem, they serve as other two components of random axial vector potential, causing a shift of Weyl nodes in the $xy$ plane. The second intranode term describes a random axial chemical potential, and at the microscopic level it originates from the random Doppler shift. Therefore, the RG analysis has to be carried out by including these additional scattering processes. We will assume the $V_{j,45}$'s and $V_{45}$ to be independent random variables following Gaussian white noise distributions, respectively possessing the coupling constants $\Delta_{1,45}=\Delta_{2,45}=\Delta_3$ and $\Delta_4$. Therefore, we perform the RG analysis of the following replicated effective action
\begin{widetext}
\begin{eqnarray}
&&\bar{S}_{MW}=\int d^3x d\tau \Psi^\dagger_a [\partial_\tau - i v \partial_j \Gamma_j]\Psi_a-\frac{\Delta_1}{2} \int d^3x d\tau d\tau^\prime [\Psi^\dagger_a \Gamma_3\Gamma_{45} \Psi_a](\mathbf{x},\tau)[\Psi^\dagger_b \Gamma_3 \Gamma_{45}\Psi_b](\mathbf{x},\tau^\prime)-\frac{\Delta_2}{2} \int d^3x d\tau d\tau^\prime \nonumber \\ && \times \{[\Psi^\dagger_a \Gamma_4 \Psi_a](\mathbf{x},\tau)[\Psi^\dagger_b \Gamma_4 \Psi_b](\mathbf{x},\tau^\prime)+[\Psi^\dagger_a \Gamma_5 \Psi_a](\mathbf{x},\tau)[\Psi^\dagger_b \Gamma_5 \Psi_b](\mathbf{x},\tau^\prime)\}-\frac{\Delta_3}{2} \sum_{j=1}^{2} \int d^3x d\tau d\tau^\prime [\Psi^\dagger_a \Gamma_j\Gamma_{45} \Psi_a](\mathbf{x},\tau) \times \nonumber \\ && [\Psi^\dagger_b \Gamma_j \Gamma_{45} \Psi_b](\mathbf{x},\tau^\prime)-\frac{\Delta_4}{2} \int d^3x d\tau d\tau^\prime [\Psi^\dagger_a \Gamma_{45} \Psi_a](\mathbf{x},\tau)[\Psi^\dagger_b \Gamma_{45}\Psi_b](\mathbf{x},\tau^\prime).
\end{eqnarray}
\end{widetext}
Under the scale transformation $\mathbf{x} \to \mathbf{x} e^l$, $\tau \to \tau e^{l}$, $\Psi \to \Psi e^{-3l/2}$ the quadratic part of the action remains invariant. But, the disorder couplings change according to $\Delta_j (l)=\Delta_j(0) e^{-l}$. Therefore, weak disorder is an irrelevant perturbation. After carrying out a one loop RG calculation (this is controlled by a $d=2+\epsilon$ continuation) we find the following RG flow equations
\begin{eqnarray}
\frac{d\Delta_1}{dl}&=&\Delta_1\left[-1-\frac{2\Delta_1}{3}-\frac{4\Delta_2}{3}+\frac{4\Delta_3}{3}-\frac{2\Delta_4}{3}\right]+\frac{4\Delta^2_2}{3}\nonumber \\ &&+\frac{8\Delta_3\Delta_4}{3}, \\
\frac{d\Delta_2}{dl}&=&\Delta_2\left[-1-\frac{2\Delta_1}{3}-\frac{4\Delta_3}{3}+2\Delta_4\right],\\
\frac{d\Delta_3}{dl}&=&\Delta_3\left[-1+\frac{2\Delta_1}{3}-\frac{4\Delta_2}{3}+\frac{2\Delta_4}{3}\right] +\frac{4\Delta^2_2}{3}\nonumber \\ &&+\frac{4\Delta_1\Delta_4}{3},\\
\frac{d\Delta_4}{dl}&=&\Delta_4\left[-1+2\Delta_1-4\Delta_2+4\Delta_3+2\Delta_4\right]+\frac{4\Delta^2_3}{3} \nonumber \\ &&+\frac{8\Delta_1\Delta_3}{3},
\end{eqnarray}
and a scale dependent dynamic scaling exponent $z(l)=1+\Delta_1(l)+2\Delta_2(l)+2\Delta_3(l)+\Delta_{4}(l)$. Apart from the attractive clean fixed point, these flow equations support a repulsive fixed point at $\Delta_1=\Delta_2=\Delta_3=0$ and $\Delta_4=1/2$, with a dynamic scaling exponent $z=3/2$. \emph{Therefore, the universality class of the semimetal to diffusive metal transition of MW fermions is governed by the random axial chemical potential}. This claim can be further substantiated by numerically solving the coupled RG flow equations, with the initial condition $\Delta_1(0)=\Delta_2(0) \neq 0$ and $\Delta_3(0)=\Delta_4(0)=0$. Notice that the intranode part of the random chemical potential disorder by itself is an irrelevant perturbation, and the interplay of intranode and internode scattering processes is essential for driving the phase transition. In Fig.~\ref{RGFlow} (a) we plot the disorder couplings $\Delta_j(l)$when $\Delta_1(0)=\Delta_2(0)=0.8701$. At a large length scale, all the renormalized disorder couplings flow to zero, thus indicating a perturbatively stable ThSM2 phase. By contrast, for $\Delta_1(0)=\Delta_2(0)=0.8702$, renormalized axial chemical potential coupling $\Delta_4$ flows to strong coupling, as shown in Fig.~\ref{RGFlow} (b). This signals a disorder driven diffusive metal phase. Notice that the renormalized strengths of the other three  disorder couplings are negligible in comparison to $\Delta_4$, which helps us to verify that the universality class of disorder driven semimetal to metal transition is indeed controlled by random axial chemical potential.  

For a single Weyl cone with a random axial chemical potential, one can directly apply the arguments for non-perturbative effects of rare regions as in Ref.~\cite{Nandkishore2014} to show the existence of low energy quasi-localized eigenstates that produce a non-zero but exponentially small DOS at weak disorder. The fact that the universality class of the transition for MW fermions reduces to that of the random axial chemical potential provides a plausible explanation for why the non-perturbative effects of disorder (described in detail in the next section) in this particle hole symmetric model are well described by rare regions in the presence of potential disorder, which breaks this symmetry. Essentially, nonperturbative rare region effects are outside the realm of the perturbative RG theory developed in this section (and the situation does not change in higher loops in RG either) with the very basic ingredient of the RG argument, namely, that weak disorder is irrelevant (which also remains true in the self-consistent Born approximation) breaks down in the presence of disorder-induced rare region effects leading to a destabilization of the ThSM phase to the ThDM phase already at infinitesimal disorder.  We study this nonperturbative physics numerically in Sec.~\ref{sec:non-pert}.

\section{Thermal diffusive metal}
\label{sec:ThermalDiffMetal}
In this appendix we briefly study the properties of the low energy DOS deep in the ThDM phase.
In order to make sure we are not in an insulating phase, we consider both the average and typical DOS.
Both of these quantities are nonzero, and the DOS follows closely the non-linear sigma model analysis~\cite{Senthil1998,*Senthil1999,*Senthil1999a,Bocquet2000}.
\begin{figure}[!ht]
\includegraphics[width=\columnwidth]{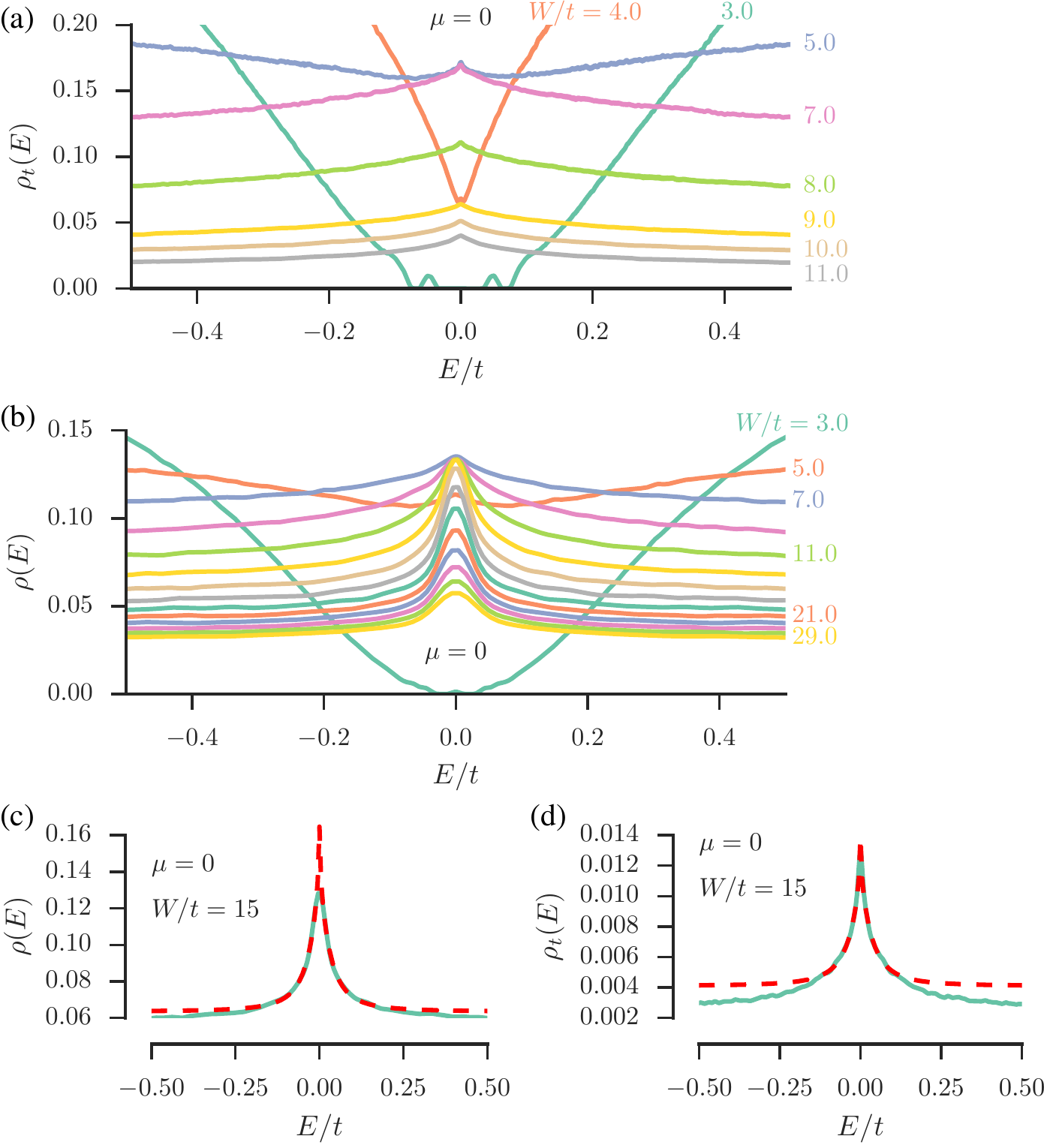}
  \caption{The peak in both the normal-DOS (a) and typical-DOS (b) around $E=0$ as predicted by perturbation theory for a ThDM. 
The peak becomes well-defined at finite disorder and persists up to all disorders considered here. 
The DOS is calculated at $L=60$ and the typical-DOS is calculated at $L=30$.
 (c) We show here two example fits to the expected analytical form for the anti-localizaiton peak in $\rho(E)$ to the numerical data defined with the integral in Eq.~\eqref{eq:DOSpeak} for small $E$. (d)
    The fit (the dashed red line) works surprisingly well to even lower energies for $\rho_t(E)$
\label{fig:peaks}} 
\end{figure}
As predicted, we see a characteristic peak in the average and typical DOS within this regime as seen in Fig.~\ref{fig:peaks} (a) and (b).
In particular, from perturbation theory quantum interference phenomena induces a peak in the DOS that can be calculated from
\begin{align}
  \rho(E) = \bar{\rho} + \frac1\pi\Re \int \frac{d^3 k}{(2\pi)^3} \frac{1}{D k^2 - 2 i E}. \label{eq:DOSpeak}
\end{align}
where $\bar \rho$ is some constant, $\Re$ is specifies the real part, and $D$ is the diffusion constant (we fit our peak to $\bar \rho$, $D$, and a cutoff scale $\Lambda$).
In particular, near $E=0$, we have 
\begin{align}
  \rho(E)- \rho(0) \sim - \sqrt{|E|},
\end{align}
with $\rho(0) >0$.
The data is well fit by the analytic form as indicated in Fig.~\ref{fig:peaks}(c).
It is important to note that this peak does not manifest itself until well after the avoided QCP and its onset is indicated by a dashed green line in Fig.~\ref{fig:basicphasediagram}.
In fact, the fit is even better for the typical DOS as seen in Fig.~\ref{fig:peaks}(d).
However, higher energy states begin to localize as indicated by the typical DOS and we eventually find a transition into an Anderson insulator.

\section{Estimating cross over and phase boundaries}
\label{sec:AppendixA}
\subsection{The avoided QCP between ThSM to ThDM}
For the ThSM$n$ to ThDM crossover, we have studied $\mu/t = 0,\, 2.32,\, 2.42,\, 2.52,\, 5.0$ at various system sizes ranging from $L=60-140$ in steps of $20$. As a function of $W$, we analyze $\rho(0)$ for various system sizes to obtain an estimate of 
the avoided critical line $W_c(\mu)$ between the ThSM$n$ and ThDM crossovers. 
As $1/L \rightarrow 0$, $\rho(0)$ tends to its rare-region value $\rho(0)\ll 1$ in the ThSM$n$ regime, but in the ThDM regime, it saturates to a larger value $\rho(0)\sim O(1)$, and for $\mu/t = 0,\, 2.32,\, 2.42,\, 2.52$ we use this to locate $W_c(\mu)$, see Fig.~\ref{fig:mu0SMtoDM}.
However, in general tracking the size dependence is complicated by the renormalization of $\mu$: The number of states at $E=0$ can fluctuate at finite disorder, and with the numerics presented in this section we cannot distinguish between rare-region effects and numerical background effects when the zero-energy state is not present.
We therefore cannot use this approach in general.
For the other values of $\mu$ along this transition line ($\mu = 0-10$ in increments of $0.5$) we focus on $L=120$ and fit a power law well into the ThDM regime to find the transition point, i.e.\ from $\rho(0) \propto {(W - W_c)}^b$ on the ThDM side of the transition for $L=120$.
From this we obtain the approximate location of the avoided quantum critical point between the ThSM and ThDM regimes as shown in Fig.~\ref{fig:basicphasediagram}. 
We have checked that these two procedures give consistent estimates of $W_c(\mu)$.

\subsection{ThSM4 to ThSM2}
For the crossover between the ThSM4 to ThSM2 regimes we utilize the fact that for a fixed $L$, $\rho(0)$ has a maximum at the transition for $W=0$ (see Fig~\ref{fig:cleanPhaseDiagram}(c)).
We can understand this maximum by an increase in the number of zero energy states due to the three anisotropic points, and we can easily track it as seen in Fig.~\ref{fig:rho0_W2_thSMthSM}. 
Due to the perturbative irrelevance of disorder to the anisotropic Weyl nodes and an analysis of $\rho(E)$ (see Section~\ref{sec:thsm4-thsm2}) we conclude that this is a signature of the crossover. This maximum broadens and eventually intersects the avoided quantum critical line.

\subsection{ThSM2/ThDM to ThBI}
To determine the transition from the ThDM to the ThBI phase, the value of $\rho(0)$ drops abruptly (due to an average gap there are very few if not zero states at or near $E=0$, even at finite size) which is easily discernible up to $W\sim 5 t$. Since $\rho(0)$ isn't an order parameter for the transition we use two complementary methods to find the critical line $\mu_I(W)$: (A) By fitting $\rho(E) \sim E^{d/z^*-1}$ to the DOS inside the ThSM2 
regime and observing where $z^*\ll 1$ (numerically $d/z^*\sim O(10^2)$ suddenly, see Fig.~\ref{fig:TransitiontoBIW7}) as a result of an average gap in the spectrum.
(B) Approaching the transition from the ThBI, we find the average band gap in the ThBI phase and extrapolate it to zero via a power law fit, see Fig.~\ref{fig:DOSW6plots}(a).

\bibliography{reference1}

\end{document}